\documentclass[prd,a4paper,showpacs,preprintnumbers,nofootinbib,amsmath]{revtex4} 

\usepackage{graphicx}
\usepackage{subfigure}

\begin{document} 

\title{Dilepton production from polarized hadron hadron collisions}

\author{S.~Arnold$^a$, A.~Metz$^b$, M.~Schlegel$^c$ \\[0.1cm]
$^a${\it Institut f\"ur Theoretische Physik II, Ruhr-Universit\"at Bochum,
 44780 Bochum, Germany} \\
$^b${\it Department of Physics, Temple University, Philadelphia, 
 PA 19122-6082, U.S.A.} \\
$^c${\it Theory Center, Jefferson Lab, 12000 Jefferson Avenue,
 Newport News, VA 23606, U.S.A.}}

\date{\today}

\begin{abstract}
In this paper we present a comprehensive formalism for dilepton production from the 
collision of two polarized spin-$\tfrac{1}{2}$ hadrons by identifying the general
angular distribution of the cross section in combination with a complete set of 
structure functions.
The various structure functions are computed in the parton model approximation 
where we mainly consider the case when the transverse momentum of the dilepton 
pair is much smaller than its invariant mass.
In this kinematical region dilepton production can be described in terms of
transverse momentum dependent parton distributions.
\end{abstract}



\maketitle

\section{Introduction}

During the past decades dilepton production in high-energy hadron hadron 
collisions (the so-called Drell-Yan (DY) process~\cite{drell_70a,drell_70b}) 
has played an important role in order to pin down parton distributions (PDFs) 
of hadrons.
While the main focus was on PDFs of the nucleon, also information on the 
partonic structure of the pion was already obtained through Drell-Yan 
measurements.
The crucial tool required for the extraction of PDFs is the QCD-factorization 
theorem~\cite{bodwin_84,collins_85,collins_88,collins_89} which applies if 
the invariant mass of the dilepton pair is sufficiently large. 
Experimentally, the Drell-Yan process is quite challenging because of the 
relatively low counting rates.
On the other hand, from the theoretical point of view it is the cleanest 
hard hadron hadron scattering process.
The fact that no hadron is detected in the final state simplifies the proof
of factorization in comparison to hadron hadron collisions with hadronic final 
states.
This important point is one of the main reasons for the continued interest in 
the Drell-Yan reaction.

Currently, not less than six programs for future Drell-Yan measurements are 
pursued.
These plans comprise dilepton production in nucleon nucleon collisions 
(at RHIC~\cite{RHIC_08}, J-PARC (KEK)~\cite{JPARC_06,JPARC_07}, 
IHEP (Protvino)~\cite{abramov_05}, and at the JINR (Dubna)~\cite{sissakian_08}), 
in antiproton nucleon collisions 
(at FAIR (GSI)~\cite{PAX_05}), as well as in pion nucleon collisions 
(at COMPASS (CERN)~\cite{COMPASS}).
Past measurements exclusively considered the unpolarized cross section, 
but all future programs are also aiming at polarization measurements.
Including polarization of the incoming hadrons opens up a variety of new 
opportunities for studying the strong interaction in both the perturbative and 
the nonperturbative regime.
Here we only mention the access to the transversity distribution of the 
nucleon~\cite{ralston_79,jaffe_91a,cortes_91,barone_97,martin_97,martin_99,anselmino_04a,efremov_04a}, 
and to transverse momentum dependent parton distributions (TMDs).
The TMDs not only depend on the longitudinal momentum of a parton inside a
hadron but also on its (intrinsic) transverse momentum and, in general, 
describe the strength of various intriguing spin-spin or spin-orbit 
correlations of the parton-hadron system 
(see Refs.~\cite{barone_01,mulders_95,bacchetta_06,dalesio_07} for more 
information on TMDs).

In order to analyze upcoming data from polarized Drell-Yan measurements it is 
necessary to have a general and concise formalism at hand.
The main motivation for writing the present paper is to provide such a 
framework.
To this end we decompose the hadronic tensor of the polarized Drell-Yan process 
in terms of 48 basis tensors which are multiplied by structure functions.
We limit ourselves to photon exchange and do not consider weak interaction 
effects.
To ensure electromagnetic gauge invariance of the hadronic tensor we make use
of a projector method proposed in Ref.~\cite{bardeen_68}.
On the basis of the hadronic tensor we then write down the general structure 
of the angular distribution of the Drell-Yan process.
This step is most conveniently done in a dilepton rest frame like the
Collins-Soper frame~\cite{collins_77}.
In addition to our model-independent results we also consider the process
in the parton model approximation, where we distinguish between two cases: 
(1) cross section integrated upon the transverse momentum $q_T$ of the 
dilepton pair; (2) cross section kept differential in $q_T$ and $q_T \ll q$, 
where $q$ is the invariant mass of the dilepton pair.
While in the former case one ends up with ordinary forward PDFs, in the latter 
TMDs enter in the parton model description and in a full 
QCD treatment~\cite{collins_84,ji_04a,collins_04}.

In addition to our model-independent treatment we also consider the process in 
the parton model approximation by concentrating on the situation when the 
cross section is kept differential in the transverse momentum $q_T$ of the 
dilepton pair.
In this case TMDs enter the parton model description as well as a full 
QCD treatment~\cite{collins_84,ji_04a,collins_04}.

Part of the results presented here were already given in the 
literature~\cite{ralston_79,pire_83,tangerman_94a,boer_99}, and we comment on 
other work during the course of the manuscript.
However, to the best of our knowledge, a complete formalism for the polarized 
Drell-Yan process has not been worked out before.

The manuscript is organized as follows. 
In Section II we fix part of our notation and give the general form
of the cross section in the one-photon exchange approximation.
Section III contains the decomposition of the hadronic tensor in terms
of basis tensors and structure functions, while in Section IV some discussion 
on reference frames is given.
In Section V we present the general angular distribution of the polarized 
Drell-Yan process which can be derived from the results of Section III 
in a straightforward manner.
Section VI contains the results for the structure functions in the parton
model approximation.
We conclude in Section VII.

\section{Cross section in one-photon exchange approximation}

\begin{figure}[t]
\begin{center}
\subfigure[]
{\includegraphics[scale=0.5]{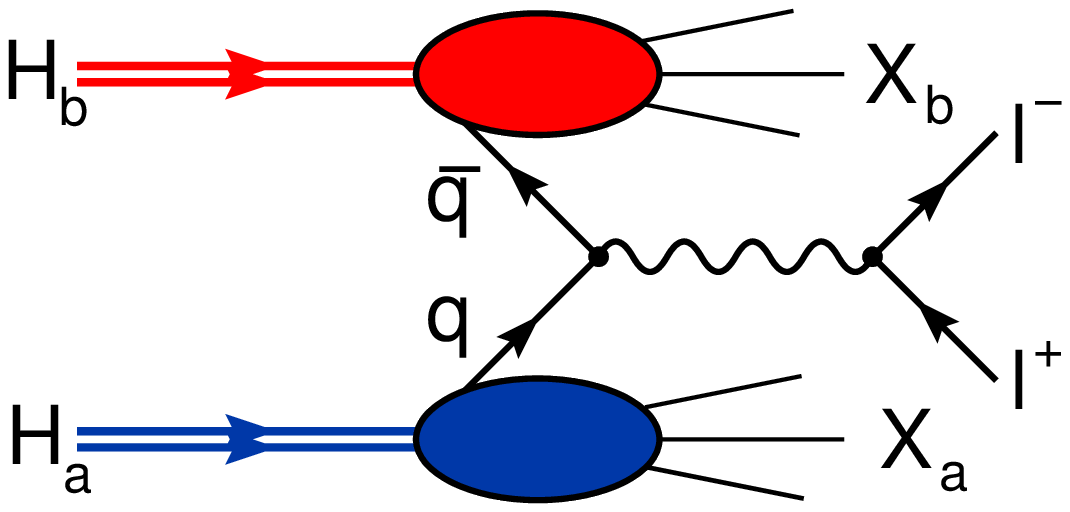}} 
\hspace{2cm}
\subfigure[]
{\includegraphics[scale=0.5]{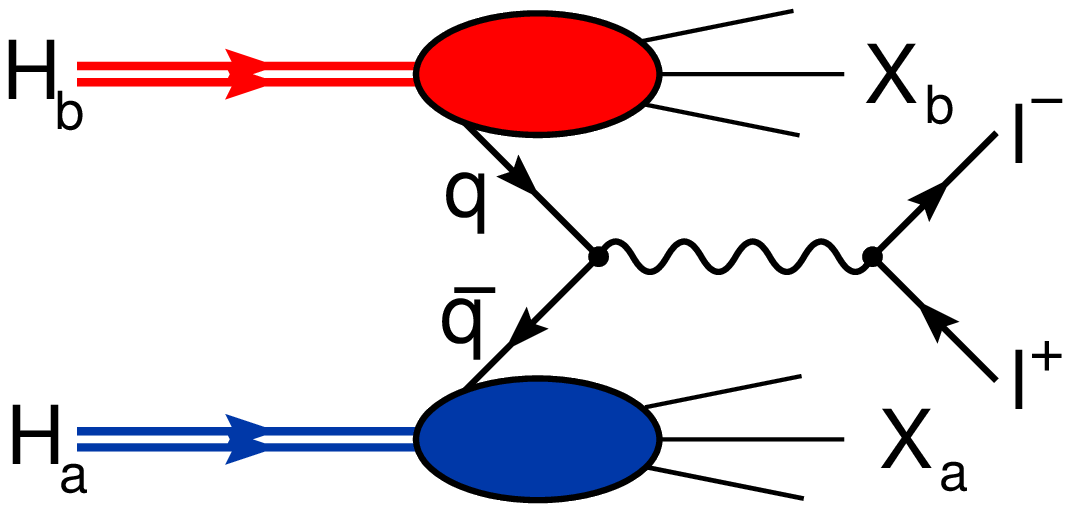}}
\end{center}
\caption{Amplitude for dilepton production in parton model approximation.
Both diagrams have to be taken into account.
The spectator systems $X_a$ and $X_b$ of the two hadrons are not detected.
\label{f:dy_amp}}
\end{figure}

To be now specific we consider the dilepton production
\begin{equation} \label{e:dilepton}
H_a(P_a,S_a) + H_b(P_b,S_b) \to l^-(l,\lambda) + l^+(l^{\prime},\lambda^{\prime}) + X \,,
\end{equation}
with $(P_a,S_a)$ and $(P_b,S_b)$ denoting the 4-momenta and the spin vectors of the 
incoming hadrons.
One has $P_a^2 = M_a^2$, $P_a \cdot S_a = 0$, $S_a^2 = -1$, and corresponding 
relations for the second hadron.
Throughout this work the mass of the leptons in the final state is neglected.
We will sum over the helicities $\lambda$, $\lambda^{\prime}$ of the leptons.
 
At large invariant mass $q$ of the dilepton pair the process~(\ref{e:dilepton}) 
can approximately be described in the Drell-Yan model~\cite{drell_70a,drell_70b},
which corresponds to the parton model approximation.
According to this approach a quark from hadron $H_a$ and an antiquark from 
hadron $H_b$ (and vice versa) annihilate into a timelike virtual photon which 
subsequently decays into a lepton pair 
(see Fig.~\ref{f:dy_amp}).\footnote{As already mentioned we do not
consider weak interaction effects.}
This means the process proceeds according to
\begin{equation} \label{e:gammadecay}
H_a + H_b \to \gamma^{\ast}(q) + X \to  l^- + l^+ + X \,,
\end{equation}
where the 4-momentum of the virtual photon is given by 
$q = l + l^{\prime}$.\footnote{In our notation the symbol $q$ describes both the 
4-momentum of the virtual photon as well as the invariant mass $\sqrt{q^2}$ of the 
dilepton pair. This should, however, not lead to any confusion.}
Note that the meaning of~(\ref{e:gammadecay}) remains valid if higher order QCD 
corrections are taken into account.

In the one-photon exchange approximation the (frame-independent) cross section 
of the Drell-Yan process is given by
\begin{equation} \label{e:xs_1}
\frac{l^0 l^{\prime 0} \, d\sigma}{d^3\vec{l} \, d^3\vec{l'}}
= \frac{\alpha_{em}^2}{F \, q^4} \, L_{\mu\nu} W^{\mu\nu} \,,
\end{equation}
where
\begin{equation}
F = 4 \sqrt{(P_a \cdot P_b)^2 - M_a^2 M_b^2} 
\end{equation}
represents the flux of the incoming hadrons.
If hadron masses are neglected one can write
$F = 2s = 2(P_a + P_b)^2$.
The fine structure constant is related to the elementary charge through
$\alpha_{em} = e^2/4\pi$.
In Eq.~(\ref{e:xs_1}) the quantity $L^{\mu\nu}$ denotes the spin-averaged 
leptonic tensor,
\begin{equation} \label{e:leptontensor}
L^{\mu\nu} = \sum_{\lambda,\lambda'} 
 \big( \bar{u}(l,\lambda) \gamma^{\mu} v(l',\lambda')  \big) 
 \big( \bar{u}(l,\lambda) \gamma^{\nu} v(l',\lambda')  \big)^{\ast}
= 4 \bigg( l^{\mu} l^{\prime\nu} + l^{\nu} l^{\prime\mu} -\frac{q^2}{2} g^{\mu\nu} \bigg) \,,
\end{equation}
while
\begin{equation}
W^{\mu\nu}(P_a,S_a; P_b,S_b; q) = 
\frac{1}{(2\pi)^4} \int d^4x \, e^{i q \cdot x} \, 
\langle P_a,S_a;P_b,S_b \,|\, J_{em}^{\mu}(0) J_{em}^{\nu}(x) \,|\, P_a,S_a;P_b,S_b \rangle
\end{equation}
is the hadronic tensor, which is determined by the electromagnetic current operator
$J_{em}^{\mu}$.

The tensor $W^{\mu\nu}$ {\it a priori} is unknown and contains the information on the
hadron structure.
It has to fulfill certain constraints due to electromagnetic gauge invariance, parity, 
and hermiticity.
In this order the constraints read
\begin{eqnarray} \label{e:gaugeinv}
q_{\mu} W^{\mu\nu}(P_a,S_a; P_b,S_b; q) & = &
 q_{\nu} W^{\mu\nu}(P_a,S_a; P_b,S_b; q) \, = \, 0 \,,
\\ \label{e:parity}
W^{\mu\nu}(P_a,S_a; P_b,S_b; q) & = &
 W_{\mu\nu}(\bar{P}_a,-\bar{S}_a; \bar{P}_b,-\bar{S}_b; \bar{q}) \,,  
\\ \label{e:hermiticity}
W^{\mu\nu}(P_a,S_a; P_b,S_b; q) & = & 
 \big[ W^{\nu\mu}(P_a,S_a; P_b,S_b; q) \big]^{\ast} \,, 
\end{eqnarray}
where the notation $\bar{v}^{\mu} = v_{\mu}$ for a generic 4-vector $v$ is used.
In Section III, by imposing the relations~(\ref{e:gaugeinv})--(\ref{e:hermiticity}), 
the hadronic tensor is decomposed into a set of 48 basis tensors multiplied by 
scalar functions (structure functions).
In doing so the conditions~(\ref{e:gaugeinv}) and~(\ref{e:parity}) considerably 
reduce the number of allowed basis tensors, while the hermiticity 
constraint~(\ref{e:hermiticity}) implies that the structure functions are real.
Note that time-reversal does not impose any constraint on the hadronic tensor,
because this operation converts the two-particle hadronic in-state into a 
two-particle out-state, and both states are not related.
In Section VI the hadronic tensor is considered in the parton model 
approximation.

The angular distribution of the Drell-Yan cross section is most conveniently be
considered in a dilepton rest frame like the Collins-Soper frame~\cite{collins_77} 
or the Gottfried-Jackson frame~\cite{gottfried_64}.
In any dilepton rest frame, one can rewrite Eq.~(\ref{e:xs_1}) according to
\begin{equation} \label{e:xs_2}
\frac{d\sigma}{d^4 q \, d\Omega}
= \frac{\alpha_{em}^2}{2\, F \, q^4} \, L_{\mu\nu} W^{\mu\nu} \,,
\end{equation}
where the solid angle $\Omega$ specifies the orientation of the leptons.
In Section IV we elaborate a bit more on reference frames with the main focus on 
the center-of-mass frame ({\it cm}-frame) and the Collins-Soper frame (CS-frame).

\section{Hadronic Tensor}

The total hadronic tensor can be decomposed into the unpolarized, single polarized 
(for hadron $H_a$ and hadron $H_b$), and double polarized tensor according to
\begin{equation} \label{e:tensor_decomp}
W^{\mu\nu} = W_u^{\mu\nu} + W_a^{\mu\nu} + W_b^{\mu\nu} + W_{ab}^{\mu\nu} \,.
\end{equation}
In the following we merely have to consider the symmetric part of $W^{\mu\nu}$
because the spin-averaged leptonic tensor in~(\ref{e:leptontensor}) is symmetric 
under the exchange $\mu \leftrightarrow \nu$.

\subsection{Unpolarized case}
Since the unpolarized tensor depends on the 4-vectors $q^{\mu}$, $P_a^{\mu}$, and
$P_b^{\mu}$ one can immediately write down the tensor basis
\begin{align} \label{e:tensorba_u_0}
h_{u,1}^{\mu\nu} & \, = \, 
g^{\mu\nu} \,,
\nonumber \displaybreak[0] \\
h_{u,2}^{\mu\nu} & \, = \,
q^{\mu}q^{\nu} \,,
\nonumber \displaybreak[0] \\
h_{u,3}^{\mu\nu} & \, = \, 
P_a^{\mu} P_a^{\nu} \,,
\nonumber \displaybreak[0] \\ 
h_{u,4}^{\mu\nu} & \, = \, 
P_b^{\mu} P_b^{\nu} \,,
\nonumber \displaybreak[0] \\ 
h_{u,5}^{\mu\nu} & \, = \, 
q^{\mu} P_a^{\nu} + q^{\nu} P_a^{\mu} \,,
\nonumber \displaybreak[0] \\ 
h_{u,6}^{\mu\nu} & \, = \, 
q^{\mu} P_b^{\nu} + q^{\nu} P_b^{\mu} \,,
\nonumber \displaybreak[0] \\
h_{u,7}^{\mu\nu} & \, = \, 
P_a^{\mu} P_b^{\nu} + P_a^{\nu} P_b^{\mu} \,.
\end{align}
The expressions in~(\ref{e:tensorba_u_0}) constitute a complete list of basis 
tensors being in accordance with the parity constraint~(\ref{e:parity}).
Therefore one can write in a first step
\begin{equation} \label{e:tensor_u_0}
W_u^{\mu\nu} = \sum_{i=1}^7 h_{u,i}^{\mu\nu} \tilde{V}_{u,i} \,,
\end{equation}
where the structure functions $\tilde{V}_{u,i}$ depend on the invariants
$P_a \cdot q$, $P_b \cdot q$, and $q^2$.

So far we have not yet used the gauge invariance 
constraint~(\ref{e:gaugeinv}) which, in fact, implies that not all of the 
$\tilde{V}_{u,i}$ are independent.
Contracting the tensor in~(\ref{e:tensor_u_0}) with the 4-momentum of the 
virtual photon and imposing~(\ref{e:gaugeinv}) one readily finds 
\begin{eqnarray} \label{e:gaugeinv_elim}
0 & = & 
 \tilde{V}_{u,1} + q^2 \, \tilde{V}_{u,2} 
 + P_a \cdot q \, \tilde{V}_{u,5} + P_b \cdot q \, \tilde{V}_{u,6} \,,
\nonumber \\
0 & = & 
 P_a \cdot q \, \tilde{V}_{u,3} + q^2 \, \tilde{V}_{u,5} 
 + P_b \cdot q \, \tilde{V}_{u,7} \,,
\nonumber \\
0 & = & 
 P_b \cdot q \, \tilde{V}_{u,4} + q^2 \, \tilde{V}_{u,6} 
 + P_a \cdot q \, \tilde{V}_{u,7} \,.
\end{eqnarray}
These three relations follow because in $W_u^{\mu\nu} q_{\nu}$ the terms 
proportional to $q^{\mu}$, $P_a^{\mu}$, and $P_b^{\mu}$ must vanish separately.
Now one can use~(\ref{e:gaugeinv_elim}) to eliminate three structure functions 
and consequently ends up with a hadronic tensor given by just four independent 
structure functions that are multiplied by four independent basis tensors.
The explicit form of the basis tensors depends of course on which of the structure 
functions are eliminated.

Though this procedure of implementing gauge invariance in principle is 
straightforward it gets rather cumbersome for single and double polarization 
because in those cases considerably more structure functions and basis tensors
are involved.
Therefore we resort to an alternative and very elegant method proposed in
Ref.~\cite{bardeen_68} which makes use of projection operators.
We define\footnote{Note that the projection operator is not unique~\cite{bardeen_68}.
One can also define an operator involving the hadron momentum $P_a$ or $P_b$.}
\begin{equation}
P^{\mu\nu} = g^{\mu\nu} - \frac{q^{\mu} q^{\nu}}{q^2} \,,
\end{equation}
and let this operator act on the basis tensors in~(\ref{e:tensorba_u_0}) 
according to
\begin{equation} \label{e:proj_appl}
P^{\mu}_{\hphantom{\mu} \rho} \, h_{u,i}^{\rho \sigma} \, P_{\sigma}^{\hphantom{\sigma} \nu} \,.
\end{equation}
Because of the property
\begin{equation}
q_{\mu} P^{\mu\nu} = P^{\mu\nu} q_{\nu} = 0
\end{equation}
the tensors in~(\ref{e:proj_appl}) vanish for $i = 2, \, 5, \, 6$, while the remaining
four nonzero tensors are gauge invariant by construction.
This means that one arrives at the following final form of the unpolarized hadronic 
tensor:
\begin{equation} \label{e:tensor_u_1}
W_u^{\mu\nu} = \sum_{i=1}^4 t_{u,i}^{\mu\nu} V_{u,i} \,,
\end{equation}
with the four structure functions $V_{u,i}$, and the tensor basis
\begin{eqnarray} \label{e:tensorba_u_1}
t_{u,1}^{\mu\nu} & = &
g^{\mu\nu} - \frac{q^{\mu} q^{\nu}}{q^2} \,,
\nonumber \\
t_{u,2}^{\mu\nu} & = &
\tilde{P}_a^{\mu} \tilde{P}_a^{\nu} \,,
\nonumber \\
t_{u,3}^{\mu\nu} & = &
\tilde{P}_b^{\mu} \tilde{P}_b^{\nu} \,,
\nonumber \\
t_{u,4}^{\mu\nu} & = &
\tilde{P}_a^{\mu} \tilde{P}_b^{\nu} + \tilde{P}_a^{\nu} \tilde{P}_b^{\mu} \,.
\end{eqnarray}
In Eq.~(\ref{e:tensorba_u_1}) we make use of the vectors
\begin{equation} \label{e:papbtilde}
\tilde{P}_a^{\mu} = P_a^{\mu} - \frac{P_a \cdot q \, q^{\mu}}{q^2} \,, \qquad
\tilde{P}_b^{\mu} = P_b^{\mu} - \frac{P_b \cdot q \, q^{\mu}}{q^2} \,, 
\end{equation}
which vanish upon contraction with $q$.
Needless to say that the tensor in~(\ref{e:tensor_u_1}) is frame-independent.
The contraction with the leptonic tensor can therefore be performed in any
frame.
In the context of the parton model calculation in Section VI, for instance,
this contraction is carried out in the {\it cm}-frame.
The specific form of the tensor~(\ref{e:tensor_u_1}) is by no means unique.
Other sets of basis tensors can be found in the literature 
(see, e.g., Refs.~\cite{lam_78,lam_80,ralston_79} and 
also~\cite{boer_06}), and it is straightforward to write down relations between 
different sets. 
Here we refrain from doing so because it does not give much further insight
and, in addition, is not needed for the main purpose of this paper.
We have discussed the unpolarized case in some detail in order to outline the
procedure which is used in the following two subsections that are dealing
with hadron polarization.

\subsection{Single polarized case}

Now we proceed to the case when one of the hadrons in the initial state is 
polarized.
We first consider polarization of the hadron $H_a$, and then just quote the result 
for the case when $H_b$ is polarized.
In order to construct a tensor basis we now have also the spin vector 
$S_a^{\mu}$ at our disposal --- in addition to the 4-momenta $q^{\mu}$, 
$P_a^{\mu}$, $P_b^{\mu}$.
Imposing the parity constraint~(\ref{e:parity}) one finds the following list
of tensors which are symmetric under the exchange 
$\mu \leftrightarrow \nu$:
\begin{align} \label{e:tensorba_a_0} 
h_{a,1}^{\mu\nu}, \ldots , h_{a,7}^{\mu\nu} & \, = \, 
\varepsilon^{S_a q P_a P_b} \, 
\Big\{ g^{\mu\nu} \,, \; q^{\mu}q^{\nu} \,, \;
 P_a^{\mu} P_a^{\nu} \,, \; P_b^{\mu} P_b^{\nu} \,, \;
 q^{\mu} P_a^{\nu} + q^{\nu} P_a^{\mu} \,, \;
 q^{\mu} P_b^{\nu} + q^{\nu} P_b^{\mu} \,, \;
 P_a^{\mu} P_b^{\nu} + P_a^{\nu} P_b^{\mu} \Big\} \,,
\nonumber \\
h_{a,8}^{\mu\nu}, \; h_{a,9}^{\mu\nu} & \, = \, 
 \Big\{ S_a \cdot q \,, \; S_a \cdot P_b \Big\} \,
 (\varepsilon^{\mu q P_a P_b} \, q^{\nu} +
  \varepsilon^{\nu q P_a P_b} \, q^{\mu}) \,,
\nonumber \displaybreak[0] \\
h_{a,10}^{\mu\nu}, \; h_{a,11}^{\mu\nu} & \, = \,  
 \Big\{ S_a \cdot q \,, \; S_a \cdot P_b \Big\} \,
 (\varepsilon^{\mu q P_a P_b} \, P_a^{\nu} +
  \varepsilon^{\nu q P_a P_b} \, P_a^{\mu}) \,,
\nonumber \displaybreak[0] \\
h_{a,12}^{\mu\nu}, \; h_{a,13}^{\mu\nu} & \, = \, 
 \Big\{ S_a \cdot q \,, \; S_a \cdot P_b \Big\} \,
 (\varepsilon^{\mu q P_a P_b} \, P_b^{\nu} +
  \varepsilon^{\nu q P_a P_b} \, P_b^{\mu}) \,,
\nonumber \displaybreak[0] \\
h_{a,14}^{\mu\nu} & \, = \,
\varepsilon^{\mu S_a q P_a} \, q^{\nu} + \varepsilon^{\nu S_a q P_a} \, q^{\mu} \,,
\nonumber \displaybreak[0] \\
h_{a,15}^{\mu\nu} & \, = \, 
\varepsilon^{\mu S_a q P_b} \, q^{\nu} + \varepsilon^{\nu S_a q P_b} \, q^{\mu} \,,
\nonumber \displaybreak[0] \\
h_{a,16}^{\mu\nu} & \, = \, 
\varepsilon^{\mu S_a P_a P_b} \, q^{\nu} + \varepsilon^{\nu S_a P_a P_b} \, q^{\mu} \,,  
\nonumber \displaybreak[0] \\
h_{a,17}^{\mu\nu} & \, = \, 
\varepsilon^{\mu S_a q P_a} \, P_a^{\nu} + \varepsilon^{\nu S_a q P_a} \, P_a^{\mu} \,,  
\nonumber \displaybreak[0] \\
h_{a,18}^{\mu\nu} & \, = \, 
\varepsilon^{\mu S_a q P_b} \, P_a^{\nu} + \varepsilon^{\nu S_a q P_b} \, P_a^{\mu} \,, 
\nonumber \displaybreak[0] \\
h_{a,19}^{\mu\nu} & \, = \, 
\varepsilon^{\mu S_a P_a P_b} \, P_a^{\nu} + \varepsilon^{\nu S_a P_a P_b} \, P_a^{\mu} \,,  
\nonumber \displaybreak[0] \\
h_{a,20}^{\mu\nu} & \, = \,  
\varepsilon^{\mu S_a q P_a} \, P_b^{\nu} + \varepsilon^{\nu S_a q P_a} \, P_b^{\mu} \,, 
\nonumber \displaybreak[0] \\
h_{a,21}^{\mu\nu} & \, = \, 
\varepsilon^{\mu S_a q P_b} \, P_b^{\nu} + \varepsilon^{\nu S_a q P_b} \, P_b^{\mu} \,, 
\nonumber \displaybreak[0] \\
h_{a,22}^{\mu\nu} & \, = \, 
\varepsilon^{\mu S_a P_a P_b} \, P_b^{\nu} + \varepsilon^{\nu S_a P_a P_b} \, P_b^{\mu} \,,  
\nonumber \\
h_{a,23}^{\mu\nu} & \, = \, 
\varepsilon^{\mu q P_a P_b} \, S_a^{\nu} + \varepsilon^{\nu q P_a P_b} \, S_a^{\mu} \,. 
\end{align}
To shorten the notation we have used abbreviations like
$\varepsilon^{S_a q P_a P_b} = 
\varepsilon_{\mu\nu\rho\sigma} S_a^{\mu} q^{\nu} P_a^{\rho} P_b^{\sigma}$.
Note that the hadron spin vector can only appear linearly.
It turns out that not all of the tensors $h_{a,i}^{\mu\nu}$ in~(\ref{e:tensorba_a_0})
are independent of each other.
The identity
\begin{equation} \label{e:epsid}
g^{\alpha \beta} \, \varepsilon^{\mu \nu \rho \sigma} =
g^{\mu \beta} \, \varepsilon^{\alpha \nu \rho \sigma} +
g^{\nu \beta} \, \varepsilon^{\mu \alpha \rho \sigma} +
g^{\rho \beta} \, \varepsilon^{\mu \nu \alpha \sigma} +
g^{\sigma \beta} \, \varepsilon^{\mu \nu \rho \alpha} 
\end{equation}
allows one to eliminate several out of the 23 tensors.
To be explicit one finds 10 linearly independent relations between the 
tensors in~(\ref{e:tensorba_a_0}) which may be written in the form
\begin{align} \label{e:epsrel}
2 h_{a,1}^{\mu\nu} & \, = \, 
 - h_{a,16}^{\mu\nu} + h_{a,18}^{\mu\nu} 
 - h_{a,20}^{\mu\nu} + h_{a,23}^{\mu\nu} \,,
\nonumber \\
2 h_{a,2}^{\mu\nu} & \, = \, 
 h_{a,8}^{\mu\nu} - P_b \cdot q \, h_{a,14}^{\mu\nu} 
 + P_a \cdot q \, h_{a,15}^{\mu\nu} - q^2 \, h_{a,16}^{\mu\nu} \,,
\nonumber \displaybreak[0] \\
2 h_{a,3}^{\mu\nu} & \, = \, 
 - P_a \cdot P_b \, h_{a,17}^{\mu\nu} + M_a^2 \, h_{a,18}^{\mu\nu} 
 - P_a \cdot q \, h_{a,19}^{\mu\nu} \,,
\nonumber \displaybreak[0] \\
2 h_{a,4}^{\mu\nu} & \, = \, 
 h_{a,13}^{\mu\nu} - M_b^2 \, h_{a,20}^{\mu\nu} 
 + P_a \cdot P_b \, h_{a,21}^{\mu\nu} - P_b \cdot q \, h_{a,22}^{\mu\nu} \,,
\nonumber \displaybreak[0] \\
h_{a,5}^{\mu\nu} & \, = \,  
 h_{a,10}^{\mu\nu} - P_b \cdot q \, h_{a,17}^{\mu\nu} 
 + P_a \cdot q \, h_{a,18}^{\mu\nu} - q^2 \, h_{a,19}^{\mu\nu} \,,
\nonumber \displaybreak[0] \\
h_{a,5}^{\mu\nu} & \, = \,  
 - P_a \cdot P_b \, h_{a,14}^{\mu\nu} + M_a^2 \, h_{a,15}^{\mu\nu} 
 - P_a \cdot q \, h_{a,16}^{\mu\nu} \,,
\nonumber \displaybreak[0] \\
h_{a,6}^{\mu\nu} & \, = \,  
 h_{a,9}^{\mu\nu} - M_b^2 \, h_{a,14}^{\mu\nu} 
 + P_a \cdot P_b \, h_{a,15}^{\mu\nu} - P_b \cdot q \, h_{a,16}^{\mu\nu} \,,
\nonumber \displaybreak[0] \\
h_{a,6}^{\mu\nu} & \, = \,  
 h_{a,12}^{\mu\nu} - P_b \cdot q \, h_{a,20}^{\mu\nu} 
 + P_a \cdot q \, h_{a,21}^{\mu\nu} - q^2 \, h_{a,22}^{\mu\nu} \,,
\nonumber \displaybreak[0] \\
h_{a,7}^{\mu\nu} & \, = \,  
 h_{a,11}^{\mu\nu} - M_b^2 \, h_{a,17}^{\mu\nu} 
 + P_a \cdot P_b \, h_{a,18}^{\mu\nu} - P_b \cdot q \, h_{a,19}^{\mu\nu} \,,
\nonumber \\ 
h_{a,7}^{\mu\nu} & \, = \, 
 - P_a \cdot P_b \, h_{a,20}^{\mu\nu} + M_a^2 \, h_{a,21}^{\mu\nu} 
 - P_a \cdot q \, h_{a,22}^{\mu\nu} \,.
\end{align}
On the basis of the relations in~(\ref{e:epsrel}) we choose to eliminate the tensors
$h_{a,14}^{\mu\nu}, \ldots , h_{a,23}^{\mu\nu}$.

Following Eq.~(\ref{e:proj_appl}) the projection operator $P^{\mu\nu}$ is now applied 
to the remaining tensors in order to implement electromagnetic gauge invariance.
This procedure provides, in a straightforward manner, the final form of the hadronic 
tensor for the case of single hadron polarization.
One finds
\begin{equation}
W_a^{\mu\nu} = \sum_{i=1}^8 t_{a,i}^{\mu\nu} V_{a,i} \,,
\end{equation}
with the eight structure functions $V_{a,i}$, and the tensor basis
\begin{eqnarray} \label{e:tensorba_a_1}
t_{a,1}^{\mu\nu}, \ldots , t_{a,4}^{\mu\nu} & = &
\varepsilon^{S_a q P_a P_b} \,
\bigg\{ g^{\mu\nu} - \frac{q^{\mu} q^{\nu}}{q^2} \,, \;
 \tilde{P}_a^{\mu} \tilde{P}_a^{\nu} \,, \;
 \tilde{P}_b^{\mu} \tilde{P}_b^{\nu} \,, \;
 \tilde{P}_a^{\mu} \tilde{P}_b^{\nu} + \tilde{P}_a^{\nu} \tilde{P}_b^{\mu} \bigg\} \,,
\nonumber \\
t_{a,5}^{\mu\nu}, \; t_{a,6}^{\mu\nu} & = &
 \bigg\{ S_a \cdot q \,, \; S_a \cdot P_b \bigg\} \, 
 (\varepsilon^{\mu q P_a P_b} \, \tilde{P}_a^{\nu} +
  \varepsilon^{\nu q P_a P_b} \, \tilde{P}_a^{\mu}) \,,
\nonumber \\
t_{a,7}^{\mu\nu}, \; t_{a,8}^{\mu\nu} & = &
 \bigg\{ S_a \cdot q \,, \; S_a \cdot P_b \bigg\} \, 
 (\varepsilon^{\mu q P_a P_b} \, \tilde{P}_b^{\nu} +
  \varepsilon^{\nu q P_a P_b} \, \tilde{P}_b^{\mu}) \,.
\end{eqnarray}
Here we used the 4-vectors $\tilde{P}_a^{\mu}$ and $\tilde{P}_b^{\mu}$ as given 
in~(\ref{e:papbtilde}).
Note that the first four tensors in~(\ref{e:tensorba_a_1}) correspond to the four 
tensors in~(\ref{e:tensorba_u_1}) for the unpolarized case, multiplied by the 
structure $\varepsilon^{S_a q P_a P_b}$. 
It is worthwhile pointing out the following: we have chosen to first remove 
redundant tensors in~(\ref{e:tensorba_a_0}) by means of the 
identity~(\ref{e:epsid}) and then implemented gauge invariance.
If one reverses these two steps one can obtain the same final result for 
the hadronic tensor.

If the hadron $H_b$ is polarized one can now write immediately
\begin{equation}
W_b^{\mu\nu} = \sum_{i=1}^8 t_{b,i}^{\mu\nu} V_{b,i} \,,
\end{equation}
with the eight structure functions $V_{b,i}$, and the tensor basis
\begin{eqnarray} \label{e:tensorba_b_1}
t_{b,1}^{\mu\nu}, \ldots , t_{b,4}^{\mu\nu} & = &
\varepsilon^{S_b q P_b P_a} \,
\bigg\{ g^{\mu\nu} - \frac{q^{\mu} q^{\nu}}{q^2} \,, \;
 \tilde{P}_a^{\mu} \tilde{P}_a^{\nu} \,, \;
 \tilde{P}_b^{\mu} \tilde{P}_b^{\nu} \,, \;
 \tilde{P}_a^{\mu} \tilde{P}_b^{\nu} + \tilde{P}_a^{\nu} \tilde{P}_b^{\mu} \bigg\} \,,
\nonumber \\
t_{b,5}^{\mu\nu}, \; t_{b,6}^{\mu\nu} & = &
 \bigg\{ S_b \cdot q \,, \; S_b \cdot P_a \bigg\} \, 
 (\varepsilon^{\mu q P_b P_a} \, \tilde{P}_a^{\nu} +
  \varepsilon^{\nu q P_b P_a} \, \tilde{P}_a^{\mu}) \,,
\nonumber \\
t_{b,7}^{\mu\nu}, \; t_{b,8}^{\mu\nu} & = &
 \bigg\{ S_b \cdot q \,, \; S_b \cdot P_a \bigg\} \, 
 (\varepsilon^{\mu q P_b P_a} \, \tilde{P}_b^{\nu} +
  \varepsilon^{\nu q P_b P_a} \, \tilde{P}_b^{\mu}) \,.
\end{eqnarray}

In Ref.~\cite{pire_83} the case of single hadron polarization for the Drell-Yan
process was already considered.
In that paper, however, the focus of the model-independent part was on the angular 
distribution of the cross section in the CS-frame rather than on the general form 
of the hadronic tensor.
We will discuss the angular distribution of the cross section in Section V.

\subsection{Double polarized case}

Eventually, we consider the situation when both hadrons in the initial state are 
polarized.
In that case the basis tensors depend linearly on both $S_a$ and $S_b$.
A full set of tensors respecting the parity constraint~(\ref{e:parity}) and being 
symmetric under the exchange $\mu \leftrightarrow \nu$ reads
\begin{align} \label{e:tensorba_ab_0}
h_{ab,1}^{\mu\nu}, \ldots , h_{ab,7}^{\mu\nu} & \, = \, 
S_a \cdot S_b \, 
\Big\{ g^{\mu\nu} \,, \; q^{\mu}q^{\nu} \,, \;
 P_a^{\mu} P_a^{\nu} \,, \; P_b^{\mu} P_b^{\nu} \,, \;
 q^{\mu} P_a^{\nu} + q^{\nu} P_a^{\mu} \,, \;
 q^{\mu} P_b^{\nu} + q^{\nu} P_b^{\mu} \,, 
 \nonumber \\
 & \hspace{2.9cm} P_a^{\mu} P_b^{\nu} + P_a^{\nu} P_b^{\mu} \Big\} \,,
\nonumber \\
h_{ab,8}^{\mu\nu}, \ldots , h_{ab,14}^{\mu\nu} & \, = \,
S_a \cdot q \, S_b \cdot q \, 
\Big\{ g^{\mu\nu} \,, \; q^{\mu}q^{\nu} \,, \;
 P_a^{\mu} P_a^{\nu} \,, \; P_b^{\mu} P_b^{\nu} \,, \;
 q^{\mu} P_a^{\nu} + q^{\nu} P_a^{\mu} \,, \;
 q^{\mu} P_b^{\nu} + q^{\nu} P_b^{\mu} \,, 
 \nonumber \\
 & \hspace{2.9cm} P_a^{\mu} P_b^{\nu} + P_a^{\nu} P_b^{\mu} \Big\} \,,
\nonumber \displaybreak[0] \\
h_{ab,15}^{\mu\nu}, \ldots , h_{ab,21}^{\mu\nu} & \, = \, 
S_a \cdot q \, S_b \cdot P_a \, 
\Big\{ g^{\mu\nu} \,, \; q^{\mu}q^{\nu} \,, \;
 P_a^{\mu} P_a^{\nu} \,, \; P_b^{\mu} P_b^{\nu} \,, \;
 q^{\mu} P_a^{\nu} + q^{\nu} P_a^{\mu} \,, \;
 q^{\mu} P_b^{\nu} + q^{\nu} P_b^{\mu} \,, 
 \nonumber \\
 & \hspace{2.9cm} P_a^{\mu} P_b^{\nu} + P_a^{\nu} P_b^{\mu} \Big\} \,,
\nonumber \displaybreak[0] \\
h_{ab,22}^{\mu\nu}, \ldots , h_{ab,28}^{\mu\nu} & \, = \, 
S_b \cdot q \, S_a \cdot P_b \, 
\Big\{ g^{\mu\nu} \,, \; q^{\mu}q^{\nu} \,, \;
 P_a^{\mu} P_a^{\nu} \,, \; P_b^{\mu} P_b^{\nu} \,, \;
 q^{\mu} P_a^{\nu} + q^{\nu} P_a^{\mu} \,, \;
 q^{\mu} P_b^{\nu} + q^{\nu} P_b^{\mu} \,, 
 \nonumber \\
 & \hspace{2.9cm} P_a^{\mu} P_b^{\nu} + P_a^{\nu} P_b^{\mu} \Big\} \,,
\nonumber \displaybreak[0] \\
h_{ab,29}^{\mu\nu}, \ldots , h_{ab,35}^{\mu\nu} & \, = \, 
S_a \cdot P_b \, S_b \cdot P_a \, 
\Big\{ g^{\mu\nu} \,, \; q^{\mu}q^{\nu} \,, \;
 P_a^{\mu} P_a^{\nu} \,, \; P_b^{\mu} P_b^{\nu} \,, \;
 q^{\mu} P_a^{\nu} + q^{\nu} P_a^{\mu} \,, \;
 q^{\mu} P_b^{\nu} + q^{\nu} P_b^{\mu} \,, 
 \nonumber \\
 & \hspace{2.9cm} P_a^{\mu} P_b^{\nu} + P_a^{\nu} P_b^{\mu} \Big\} \,,
\nonumber \displaybreak[0] \\
h_{ab,36}^{\mu\nu}, \ldots , h_{ab,38}^{\mu\nu} & \, = \, 
S_a \cdot q \, 
\Big\{ S_b^{\mu} q^{\nu} + S_b^{\nu} q^{\mu} \,, \;
 S_b^{\mu} P_a^{\nu} + S_b^{\nu} P_a^{\mu} \,, \;
 S_b^{\mu} P_b^{\nu} + S_b^{\nu} P_b^{\mu} \Big\} \,,
\nonumber \displaybreak[0] \\
h_{ab,39}^{\mu\nu}, \ldots , h_{ab,41}^{\mu\nu} & \, = \, 
S_b \cdot q \, 
\Big\{ S_a^{\mu} q^{\nu} + S_a^{\nu} q^{\mu} \,, \;
 S_a^{\mu} P_a^{\nu} + S_a^{\nu} P_a^{\mu} \,, \;
 S_a^{\mu} P_b^{\nu} + S_a^{\nu} P_b^{\mu} \Big\} \,,
\nonumber \displaybreak[0] \\
h_{ab,42}^{\mu\nu}, \ldots , h_{ab,44}^{\mu\nu} & \, = \,
S_a \cdot P_b \, 
\Big\{ S_b^{\mu} q^{\nu} + S_b^{\nu} q^{\mu} \,, \;
 S_b^{\mu} P_a^{\nu} + S_b^{\nu} P_a^{\mu} \,, \;
 S_b^{\mu} P_b^{\nu} + S_b^{\nu} P_b^{\mu} \Big\} \,,
\nonumber \\
h_{ab,45}^{\mu\nu}, \ldots , h_{ab,47}^{\mu\nu} & \, = \, 
S_b \cdot P_a \, 
\Big\{ S_a^{\mu} q^{\nu} + S_a^{\nu} q^{\mu} \,, \;
 S_a^{\mu} P_a^{\nu} + S_a^{\nu} P_a^{\mu} \,, \;
 S_a^{\mu} P_b^{\nu} + S_a^{\nu} P_b^{\mu} \Big\} \,,
\nonumber \\
h_{ab,48}^{\mu\nu} & \, = \, 
 S_a^{\mu} S_b^{\nu} + S_a^{\nu} S_b^{\mu} \,.
 \vphantom{\Big\{}
\end{align}
Like in the case of single hadron polarization not all 48 tensors 
in~(\ref{e:tensorba_ab_0}) are independent of each other.
An explicit relation between a certain subset of the $h_{ab,i}^{\mu\nu}$ can be 
found by means of the determinant identity~\cite{meissner_08b}
\begin{equation} \label{e:detid}
D^{\mu \alpha \beta \gamma \delta ; \nu \bar{\alpha} \bar{\beta} \bar{\gamma} \bar{\delta}}
= \left| \begin{array}{ccccc}
g^{\mu\nu} & g^{\mu\bar{\alpha}} & g^{\mu\bar{\beta}} & 
   g^{\mu\bar{\gamma}} & g^{\mu\bar{\delta}} \\
g^{\alpha\nu} & g^{\alpha\bar{\alpha}} & g^{\alpha\bar{\beta}} & 
   g^{\alpha\bar{\gamma}} & g^{\alpha\bar{\delta}} \\
g^{\beta\nu} & g^{\beta\bar{\alpha}} & g^{\beta\bar{\beta}} & 
   g^{\beta\bar{\gamma}} & g^{\beta\bar{\delta}} \\
g^{\gamma\nu} & g^{\gamma\bar{\alpha}} & g^{\gamma\bar{\beta}} & 
   g^{\gamma\bar{\gamma}} & g^{\gamma\bar{\delta}} \\
g^{\delta\nu} & g^{\delta\bar{\alpha}} & g^{\delta\bar{\beta}} & 
   g^{\delta\bar{\gamma}} & g^{\delta\bar{\delta}} \\
\end{array} \right|
= 0 \,.
\end{equation}
Equation~(\ref{e:detid}) immediately implies
\begin{equation} \label{e:detrel_a}
D_{\mu \alpha \beta \gamma \delta ; \nu \bar{\alpha} \bar{\beta} \bar{\gamma} \bar{\delta}}
 \, (S_a^{\alpha} S_b^{\bar{\alpha}} + S_a^{\bar{\alpha}} S_b^{\alpha}) \,
 q^{\beta} q^{\bar{\beta}}
 P_a^{\gamma} P_a^{\bar{\gamma}}
 P_b^{\delta} P_b^{\bar{\delta}} = 0 \,,
\end{equation}
which allows one to eliminate exactly one out of the tensors 
in~(\ref{e:tensorba_ab_0}).
For the sake of symmetry we choose to eliminate the tensor $h_{ab,48}^{\mu\nu}$.
Equation~(\ref{e:detrel_a}) implies a relation of the type
\begin{equation} \label{e:detrel_b}
\Big\{ q^2 \, \big[ (P_a \cdot P_b)^2 - M_a^2 M_b^2 \big]
 - 2 P_a \cdot P_b \, P_a \cdot q \, P_b \cdot q
 + M_a^2 \, (P_b \cdot q)^2
 + M_b^2 \, (P_a \cdot q)^2 \Big\} \, 
h_{ab,48}^{\mu\nu} = \ldots \;,
\end{equation}
where the {\it r.h.s.} of~(\ref{e:detrel_b}) is a linear combination of terms in which 
most of the $h_{ab,i}^{\mu\nu} \; (i = 1 , \ldots , 47)$ enter.
We refrain from writing down this (rather lengthy) formula explicitly as it is not 
needed for the following discussion.
We also mention that the determinant identity~(\ref{e:detid}) does not lead 
to any further relation between the $h_{ab,i}^{\mu\nu}$.

To implement gauge invariance we now apply, according to Eq.~(\ref{e:proj_appl}), 
the projection operator $P^{\mu\nu}$ to the tensors in~(\ref{e:tensorba_ab_0}).
This procedure provides, in a straightforward manner, the final form of the hadronic 
tensor for the case of polarization of both hadrons.
One finds
\begin{equation} 
W_{ab}^{\mu\nu} = \sum_{i=1}^{28} t_{ab,i}^{\mu\nu} V_{ab,i} \,,
\end{equation}
with the 28 structure functions $V_{ab,i}$, and the tensor basis
\begin{align} \label{e:tensorba_ab_1}
t_{ab,1}^{\mu\nu}, \ldots , t_{ab,4}^{\mu\nu} & \, = \, 
S_a \cdot S_b \, 
\bigg\{ g^{\mu\nu} - \frac{q^{\mu} q^{\nu}}{q^2} \,, \;
 \tilde{P}_a^{\mu} \tilde{P}_a^{\nu} \,, \;
 \tilde{P}_b^{\mu} \tilde{P}_b^{\nu} \,, \;
 \tilde{P}_a^{\mu} \tilde{P}_b^{\nu} + \tilde{P}_a^{\nu} \tilde{P}_b^{\mu} \bigg\} \,,
\nonumber \\
t_{ab,5}^{\mu\nu}, \ldots , t_{ab,8}^{\mu\nu} & \, = \, 
S_a \cdot q \, S_b \cdot q \, 
\bigg\{ g^{\mu\nu} - \frac{q^{\mu} q^{\nu}}{q^2} \,, \;
 \tilde{P}_a^{\mu} \tilde{P}_a^{\nu} \,, \;
 \tilde{P}_b^{\mu} \tilde{P}_b^{\nu} \,, \;
 \tilde{P}_a^{\mu} \tilde{P}_b^{\nu} + \tilde{P}_a^{\nu} \tilde{P}_b^{\mu} \bigg\} \,,
\nonumber \displaybreak[0] \\
t_{ab,9}^{\mu\nu}, \ldots , t_{ab,12}^{\mu\nu} & \, = \, 
S_a \cdot q \, S_b \cdot P_a \, 
\bigg\{ g^{\mu\nu} - \frac{q^{\mu} q^{\nu}}{q^2} \,, \;
 \tilde{P}_a^{\mu} \tilde{P}_a^{\nu} \,, \;
 \tilde{P}_b^{\mu} \tilde{P}_b^{\nu} \,, \;
 \tilde{P}_a^{\mu} \tilde{P}_b^{\nu} + \tilde{P}_a^{\nu} \tilde{P}_b^{\mu} \bigg\} \,,
\nonumber \displaybreak[0] \\
t_{ab,13}^{\mu\nu}, \ldots , t_{ab,16}^{\mu\nu} & \, = \, 
S_b \cdot q \, S_a \cdot P_b \, 
\bigg\{ g^{\mu\nu} - \frac{q^{\mu} q^{\nu}}{q^2} \,, \;
 \tilde{P}_a^{\mu} \tilde{P}_a^{\nu} \,, \;
 \tilde{P}_b^{\mu} \tilde{P}_b^{\nu} \,, \;
 \tilde{P}_a^{\mu} \tilde{P}_b^{\nu} + \tilde{P}_a^{\nu} \tilde{P}_b^{\mu} \bigg\} \,,
\nonumber \displaybreak[0] \\
t_{ab,17}^{\mu\nu}, \ldots , t_{ab,20}^{\mu\nu} & \, = \, 
S_a \cdot P_b \, S_b \cdot P_a \, 
\bigg\{ g^{\mu\nu} - \frac{q^{\mu} q^{\nu}}{q^2} \,, \;
 \tilde{P}_a^{\mu} \tilde{P}_a^{\nu} \,, \;
 \tilde{P}_b^{\mu} \tilde{P}_b^{\nu} \,, \;
 \tilde{P}_a^{\mu} \tilde{P}_b^{\nu} + \tilde{P}_a^{\nu} \tilde{P}_b^{\mu} \bigg\} \,,
\nonumber \displaybreak[0] \\
t_{ab,21}^{\mu\nu}, \; t_{ab,22}^{\mu\nu} & \, = \, 
S_a \cdot q \, 
\bigg\{ \tilde{S}_b^{\mu} \tilde{P}_a^{\nu} + \tilde{S}_b^{\nu} \tilde{P}_a^{\mu} \,, \;
 \tilde{S}_b^{\mu} \tilde{P}_b^{\nu} + \tilde{S}_b^{\nu} \tilde{P}_b^{\mu} \bigg\} \,,
\nonumber \displaybreak[0] \\
t_{ab,23}^{\mu\nu}, \; t_{ab,24}^{\mu\nu} & \, = \, 
S_b \cdot q \, 
\bigg\{ \tilde{S}_a^{\mu} \tilde{P}_a^{\nu} + \tilde{S}_a^{\nu} \tilde{P}_a^{\mu} \,, \;
 \tilde{S}_a^{\mu} \tilde{P}_b^{\nu} + \tilde{S}_a^{\nu} \tilde{P}_b^{\mu} \bigg\} \,,
\nonumber \displaybreak[0] \\
t_{ab,25}^{\mu\nu}, \; t_{ab,26}^{\mu\nu} & \, = \, 
S_a \cdot P_b \, 
\bigg\{ \tilde{S}_b^{\mu} \tilde{P}_a^{\nu} + \tilde{S}_b^{\nu} \tilde{P}_a^{\mu} \,, \;
 \tilde{S}_b^{\mu} \tilde{P}_b^{\nu} + \tilde{S}_b^{\nu} \tilde{P}_b^{\mu} \bigg\} \,,
\nonumber \\
t_{ab,27}^{\mu\nu}, \; t_{ab,28}^{\mu\nu} & \, = \, 
S_b \cdot P_a \, 
\bigg\{ \tilde{S}_a^{\mu} \tilde{P}_a^{\nu} + \tilde{S}_a^{\nu} \tilde{P}_a^{\mu} \,, \;
 \tilde{S}_a^{\mu} \tilde{P}_b^{\nu} + \tilde{S}_a^{\nu} \tilde{P}_b^{\mu} \bigg\} \,.
\end{align}
Here we used the 4-vectors $\tilde{P}_a^{\mu}$ and $\tilde{P}_b^{\mu}$ as given 
in~(\ref{e:papbtilde}).
The vectors $\tilde{S}_a^{\mu}$ and $\tilde{S}_b^{\mu}$ are defined accordingly, i.e.,
\begin{equation} \label{e:sasbtilde}
\tilde{S}_a^{\mu} = S_a^{\mu} - \frac{S_a \cdot q \, q^{\mu}}{q^2} \,, \qquad
\tilde{S}_b^{\mu} = S_b^{\mu} - \frac{S_b \cdot q \, q^{\mu}}{q^2} \,. 
\end{equation}
Note that the first 20 tensors in~(\ref{e:tensorba_ab_1}) correspond to the four 
tensors in~(\ref{e:tensorba_u_1}) for the unpolarized case, multiplied by  
certain scalar products containing the spin vectors of the hadrons.
We again emphasize the crucial importance of the relation~(\ref{e:detrel_a}).
Without this identity the final form of the hadronic tensor would have 29 rather
than 28 basis elements.

To the best of our knowledge the general structure of the hadronic tensor for the
double polarized Drell-Yan process is a new result.
Though the double polarized case was already investigated in Ref.~\cite{ralston_79}, 
this was only done for the specific cases $q_T = 0$ and cross section integrated 
upon $q_T$.
In those cases seven basis tensors can be identified.

\subsection{Identical hadrons}

If both hadrons in the initial state are identical --- as is the case, e.g., for 
proton-proton DY --- the total hadronic tensor in Eq.~(\ref{e:tensor_decomp}) has to 
satisfy the symmetry relation
\begin{equation}
W^{\mu\nu}(P_a,S_a; P_b,S_b; q) = W^{\mu\nu}(P_b,S_b; P_a,S_a; q) \,.  
\end{equation}
This immediately implies that eight out of the 48 structure functions are symmetric 
when exchanging the momenta $P_a$ and $P_b$,
\begin{equation} \label{e:sym_v}
\begin{array}{lll}
V_{u,1}(b,a) = V_{u,1}(a,b) \,, \quad & 
V_{u,4}(b,a) = V_{u,4}(a,b) \,, \quad &
\hphantom{a} 
\\[0.2cm]
V_{ab,1}(b,a) = V_{ab,1}(a,b) \,, \quad & 
V_{ab,4}(b,a) = V_{ab,4}(a,b) \,, \quad &
V_{ab,5}(b,a) = V_{ab,5}(a,b) \,, 
\\[0.2cm]
V_{ab,8}(b,a) = V_{ab,8}(a,b) \,, \quad &
V_{ab,17}(b,a) = V_{ab,17}(a,b) \,, \quad &
V_{ab,20}(b,a) = V_{ab,20}(a,b) \,, 
\end{array}
\end{equation}
where, e.g., the first relation in~(\ref{e:sym_v}) is a shorthand of
\begin{equation}
V_{u,1}(P_b \cdot q, \, P_a \cdot q, \, q^2) = 
V_{u,1}(P_a \cdot q, \, P_b \cdot q, \, q^2) \,.
\end{equation}
Because of the symmetry property it is sufficient to know the structure functions 
in~(\ref{e:sym_v}) for just half of the allowed parameter space.
The remaining 40 structure functions fulfil the relations 
\begin{equation} \label{e:rel_v}
\begin{array}{lll}
V_{u,3}(b,a) = V_{u,2}(a,b) \,, &
\hphantom{a} & 
\hphantom{a} 
\\[0.2cm]
V_{b,1}(b,a) = V_{a,1}(a,b) \,, \quad &
V_{b,2}(b,a) = V_{a,3}(a,b) \,, \quad &
V_{b,3}(b,a) = V_{a,2}(a,b) \,, 
\\[0.2cm]
V_{b,4}(b,a) = V_{a,4}(a,b) \,, \quad &
V_{b,5}(b,a) = V_{a,7}(a,b) \,, \quad &
V_{b,6}(b,a) = V_{a,8}(a,b) \,, 
\displaybreak[0] \\[0.2cm]
V_{b,7}(b,a) = V_{a,5}(a,b) \,, \quad &
V_{b,8}(b,a) = V_{a,6}(a,b) \,, \quad &
\hphantom{a} 
\displaybreak[0] \\[0.2cm]
V_{ab,3}(b,a) = V_{ab,2}(a,b) \,, \quad &
V_{ab,7}(b,a) = V_{ab,6}(a,b) \,, \quad &
V_{ab,13}(b,a) = V_{ab,9}(a,b) \,, 
\displaybreak[0] \\[0.2cm]
V_{ab,14}(b,a) = V_{ab,11}(a,b) \,, \quad &
V_{ab,15}(b,a) = V_{ab,10}(a,b) \,, \quad &
V_{ab,16}(b,a) = V_{ab,12}(a,b) \,, 
\displaybreak[0] \\[0.2cm]
V_{ab,19}(b,a) = V_{ab,18}(a,b) \,, \quad &
V_{ab,23}(b,a) = V_{ab,22}(a,b) \,, \quad &
V_{ab,24}(b,a) = V_{ab,21}(a,b) \,, 
\\[0.2cm]
V_{ab,27}(b,a) = V_{ab,26}(a,b) \,, \quad &
V_{ab,28}(b,a) = V_{ab,25}(a,b) \,. \quad &
\hphantom{a}
\end{array}
\end{equation}
For instance the first relation in~(\ref{e:rel_v}) implies that if one knows the
structure function $V_{u,2}$ for the entire parameter space one also knows 
$V_{u,3}$.

\section{Reference Frames}

So far our treatment is frame-independent.
If, however, one wants to write down the general form of the angular distribution 
of the cross section --- as we are going to do in Section V --- one has to specify 
the reference frame.
Moreover, the parton model calculation of the hadronic tensor, carried out in 
Section VI, is naturally performed in the {\it cm}-frame.
Therefore, in the following we will consider both the {\it cm}-frame and the 
CS-frame~\cite{collins_77}, which is a particular dilepton rest frame.
In general, the angular distribution of the cross section is most conveniently
given in a dilepton rest frame.

In the {\it cm}-frame the 4-momenta $P_a^{\mu}$, $P_b^{\mu}$, and $q^{\mu}$ take the 
form
\begin{eqnarray} \label{e:pacm}
P_{a,CM}^{\mu} & = & 
 \big( P_{a,CM}^0, \, 0, \, 0, \, P_{a,CM}^3 \big) \approx
 \frac{\sqrt{s}}{2} \big( 1, \, 0, \, 0, \, 1 \big) \,,
\\ \label{e:pbcm}
P_{b,CM}^{\mu} & = & 
 \big( P_{b,CM}^0, \, 0, \, 0, \, P_{b,CM}^3 \big) \approx
 \frac{\sqrt{s}}{2} \big( 1, \, 0, \, 0, \, -1 \big) \,,
\\ \label{e:qcm}
q_{CM}^{\mu} & = &
 \big( q_{0,CM}, \, q_{T,CM}, \, 0, \, q_{L,CM} \big)
 \vphantom{\frac{\sqrt{s}}{2}} \,,
\end{eqnarray}
where the simple relation between the hadron momenta and $\sqrt{s}$ holds if 
the hadron masses are neglected.
Note that without loss of generality the transverse part of the photon momentum
is pointing into the $x$-direction.
To shorten the notation we will use $q_T \equiv q_{T,CM}$ in the following.
Equations~(\ref{e:pacm})--(\ref{e:qcm}) fix the axes of the {\it cm}-frame 
according to
\begin{equation}
\hat{e}_{x,CM} = \frac{\vec{q}_T}{q_T} \,, \qquad
\hat{e}_{y,CM} = \hat{e}_{z,CM} \times \hat{e}_{x,CM} \,, \qquad
\hat{e}_{z,CM} = \frac{\vec{P}_{a,CM}}{|\vec{P}_{a,CM}|} \,.
\end{equation}

To make the transition from the {\it cm}-frame to the CS-frame one can apply
two subsequent Lorentz boosts~\cite{collins_77}.
In a first step one boosts along the $z$-axis such that the virtual photon
no longer has a longitudinal momentum component.
In a second step one boosts along the $x$-axis such that also the transverse
momentum of the virtual photon disappears.
This leads to the following transformation matrix between the two frames:
\begin{equation} \label{e:lt}
B^{\mu}_{\;\,\nu} = \frac{1}{q} \left( 
 \begin{array}{cccc}
 q_{0,CM} \; & - \rho \, q \; & 0 \; & - q_{L,CM} \vphantom{\Big(} \\ 
 - \sin \alpha \, q_{0,CM} \; & (\cos \alpha)^{-1} \, q \; & 0  \; & 
 \sin \alpha \, q_{L,CM} \vphantom{\Big(} \\
 0 \; & 0 \; & q \; & 0   \vphantom{\Big(} \\
 - \cos \alpha \, q_{L,CM} \; & 0 \; & 0 \; & 
 \cos \alpha \, q_{0,CM} \vphantom{\Big(} 
 \end{array} \right) \,,
\end{equation}
with
\begin{equation}
\rho = \frac{q_T}{q} \,, \qquad
\sin \alpha = \frac{\rho}{\sqrt{1 + \rho^2}} \,, \qquad
\cos \alpha = \frac{1}{\sqrt{1 + \rho^2}} \,. 
\end{equation}
Applying the transformation matrix in~(\ref{e:lt}) to the 4-momenta $P_{a,CM}^{\mu}$, 
$P_{b,CM}^{\mu}$, $q_{CM}^{\mu}$ one finds, in particular, that the hadron momenta 
span the $xz$-plane.
The results are
\begin{eqnarray} \label{e:pacs}
P_{a,CS}^{\mu} & = & 
 \big( P_{a,CS}^0, \, - \sin \bar{\alpha} \, |\vec{P}_{a,CS}|, \, 
       0, \, \cos \bar{\alpha} \, |\vec{P}_{a,CS}| \big) \approx
 P_{a,CS}^0 \big( 1, \, -\sin \alpha, \, 0, \, \cos \alpha \big) \,,
\vphantom{\Big(}
\\ \label{e:pbcs}
P_{b,CS}^{\mu} & = & 
 \big( P_{b,CS}^0, \, - \sin \bar{\alpha} \, |\vec{P}_{b,CS}|, \, 
       0, \, - \cos \bar{\alpha} \, |\vec{P}_{b,CS}| \big) \approx
 P_{b,CS}^0 \big( 1, \, -\sin \alpha, \, 0, \, - \cos \alpha \big) \,,
\vphantom{\Big(}
\\ \label{e:qcs}
q_{CS}^{\mu} & = &
 \big( q, \, 0, \, 0, \, 0 \big) \,,
\vphantom{\Big(}
\end{eqnarray}
where the energies of the hadrons in the CS-frame are given by
\begin{equation}
P_{a,CS}^0 = \frac{P_a \cdot q}{q} \approx 
 \frac{\sqrt{s}}{2q} \big( q_{0,CM} - q_{L,CM} \big) \,,\qquad
P_{b,CS}^0 = \frac{P_b \cdot q}{q} \approx 
 \frac{\sqrt{s}}{2q} \big( q_{0,CM} + q_{L,CM} \big) \,.
\end{equation}
The approximate expressions in~(\ref{e:pacs}) and~(\ref{e:pbcs}) again hold if the 
hadron masses are neglected.
Note that in this case one has $\alpha = \bar{\alpha}$.
Equations~(\ref{e:pacs}), (\ref{e:pbcs}) imply that the axes in the CS-frame are fixed
by the hadron momenta according to
\begin{equation}
\hat{e}_{x,CS} = - \frac{1}{2 \,\sin \bar{\alpha}} 
 \bigg( \frac{\vec{P}_{a,CS}}{|\vec{P}_{a,CS}|} 
      + \frac{\vec{P}_{b,CS}}{|\vec{P}_{b,CS}|} \bigg)  
\,, \quad
\hat{e}_{y,CS} = \hat{e}_{z,CS} \times \hat{e}_{x,CS} \,, \quad
\hat{e}_{z,CS} = \frac{1}{2 \,\cos \bar{\alpha}} 
 \bigg( \frac{\vec{P}_{a,CS}}{|\vec{P}_{a,CS}|} 
      - \frac{\vec{P}_{b,CS}}{|\vec{P}_{b,CS}|} \bigg) \,.  
\end{equation}
In principle there are infinitely many dilepton rest frames.
Any other dilepton rest frame is related to the CS-frame through a 3-dimensional
rotation.
For instance, the frequently used Gottfried-Jackson frame~\cite{gottfried_64}, in
which the momentum of one of the hadrons is pointing into the $z$-direction, is 
connected to the CS-frame by a rotation about the $y$-axis.

One can readily invert the Lorentz transformation in~(\ref{e:lt}) and find
\begin{equation}
(B^{-1})^{\mu}_{\;\,\nu} = \frac{1}{q} \left( 
 \begin{array}{cccc}
 q_{0,CM} \; & \sin \alpha \, q_{0,CM} \; & 0 \; & 
 \cos \alpha \, q_{L,CM} \vphantom{\Big(} \\ 
 \rho \, q \; & (\cos \alpha)^{-1} \, q \; & 0 \; & 0 \vphantom{\Big(} \\
 0 \; & 0 \; & q \; & 0   \vphantom{\Big(} \\
 q_{L,CM} \; & \sin \alpha \, q_{L,CM} \; & 0 \; & 
 \cos \alpha \, q_{0,CM} \vphantom{\Big(} 
 \end{array} \right) \,.
\end{equation}
This inverse transformation is now applied to the 4-momenta of the outgoing leptons,
which in the CS-frame take the simple form
\begin{eqnarray} \label{e:lepton1_cs}
l_{CS}^{\mu} & = & \frac{q}{2} \Big( 1 , \, \sin \theta_{CS} \, \cos \phi_{CS} , \, 
                  \sin \theta_{CS} \, \sin \phi_{CS} , \, \cos \theta_{CS} \Big) \,, \\
\label{e:lepton2_cs}
l_{CS}^{\prime \mu} & = & \frac{q}{2} \Big( 1 , \, - \sin \theta_{CS} \, \cos \phi_{CS} , \, 
                  - \sin \theta_{CS} \, \sin \phi_{CS} , \, - \cos \theta_{CS} \Big) \,, 
\end{eqnarray}
i.e., the directions of both leptons are specified by the same two angles $\theta_{CS}$ 
and $\phi_{CS}$.
This feature, of course, holds in any other dilepton rest frame as well.
In the {\it cm}-frame the lepton momenta are given by
\begin{align} \label{e:lepton1_cm}
l_{CM}^{\mu} & \, = \,  \frac{1}{2} \left(
\begin{array}{c}
\big( 1 + \sin \alpha \, \sin \theta_{CS} \, \cos \phi_{CS} \big) \, q_{0,CM}
                        + \cos \alpha \, \cos \theta_{CS} \, q_{L,CM} 
\vphantom{\Big(} \\
q_T + (\cos \alpha)^{-1} \, \sin \theta_{CS} \, \cos \phi_{CS} \, q 
\vphantom{\Big(} \\
\sin \theta_{CS} \, \sin \phi_{CS} \, q 
\vphantom{\Big(} \\
\big( 1 + \sin \alpha \, \sin \theta_{CS} \, \cos \phi_{CS} \big) \, q_{L,CM}
                        + \cos \alpha \, \cos \theta_{CS} \, q_{0,CM} 
\vphantom{\Big(} 
\end{array} \right) \,,
\displaybreak[0] \\[0.2cm] \label{e:lepton2_cm}
l_{CM}^{\prime \mu} & \, = \,  \frac{1}{2} \left(
\begin{array}{c}
\big( 1 - \sin \alpha \, \sin \theta_{CS} \, \cos \phi_{CS} \big) \, q_{0,CM}
                        - \cos \alpha \, \cos \theta_{CS} \, q_{L,CM} 
\vphantom{\Big(} \\
q_T - (\cos \alpha)^{-1} \, \sin \theta_{CS} \, \cos \phi_{CS} \, q 
\vphantom{\Big(} \\
- \sin \theta_{CS} \, \sin \phi_{CS} \, q 
\vphantom{\Big(} \\
\big( 1 - \sin \alpha \, \sin \theta_{CS} \, \cos \phi_{CS} \big) \, q_{L,CM}
                        - \cos \alpha \, \cos \theta_{CS} \, q_{0,CM} 
\vphantom{\Big(} 
\end{array} \right) \,.
\end{align}
By means of these momenta one can carry out the contraction of the leptonic and 
the hadronic tensor in the {\it cm}-frame.
This is particularly convenient in connection with the parton model calculation
in Section VI.

We close this section with a brief discussion on the hadron spin vectors.
In the {\it cm}-frame one can write
\begin{eqnarray}
S_{a,CM}^{\mu} & = &
\bigg( S_{aL,CM} \, \frac{|\vec{P}_{a,CM}|}{M_a} , \,
       |\vec{S}_{aT,CM}| \, \cos \phi_{a,CM} , \,
       |\vec{S}_{aT,CM}| \, \sin \phi_{a,CM} , \,
       S_{aL,CM} \, \frac{{P}_{a,CM}^0}{M_a} \bigg) \,,
\\
S_{b,CM}^{\mu} & = &
\bigg( S_{bL,CM} \, \frac{|\vec{P}_{b,CM}|}{M_b} , \,
       |\vec{S}_{bT,CM}| \, \cos \phi_{b,CM} , \,
       |\vec{S}_{bT,CM}| \, \sin \phi_{b,CM} , \,
       - \, S_{bL,CM} \, \frac{{P}_{b,CM}^0}{M_b} \bigg) \,,
\end{eqnarray}
with the longitudinal components $S_{aL,CM}$, $S_{bL,CM}$, and the transverse 
components $\vec{S}_{aT,CM}$, $\vec{S}_{bT,CM}$.
The condition $S_a^2 = -1$ implies $(S_{aL,CM})^2 + (\vec{S}_{aT,CM})^2 = 1$
(and analogously for the hadron $H_b$).
One can also write down, e.g., $S_a^{\mu}$ in the CS-frame in terms of longitudinal
and transverse components.\footnote{The resulting expression looks a bit more 
complicated because $\vec{P}_{a,CS}$ is not pointing in the $z$-direction.}
Mainly for the following reason we prefer, however, to work with components of the
spin vectors in the {\it cm}-frame.
If one has a pure transverse polarization in the {\it cm}-frame (in the $xz$-plane),
this implies also a longitudinal polarization component in the CS-frame.
Therefore, longitudinal and transverse polarization components can get mixed
up when switching between both frames.
Since an experimental setup and also the parton model approximation have a closer 
connection to the {\it cm}-frame than to the CS-frame it is preferable to work 
with {\it cm}-frame components of the hadron spin vectors.

\section{Angular Distribution of the Cross Section}

By means of the general form of the hadronic tensor as derived in Section III one can 
now write down the full angular distribution of the DY cross section.
Since the hadronic tensor is frame-independent this can be done, in principle, for 
any reference frame.
We focus here on a dilepton rest frame because in that case the angular distribution 
takes the most compact and transparent form.
Expressing the orientation of the leptons through the CS-angles $\theta_{CS}$ and $\phi_{CS}$
(see Eqs.~(\ref{e:lepton1_cs}),~(\ref{e:lepton2_cs}), 
and~(\ref{e:lepton1_cm}),~(\ref{e:lepton2_cm})) and contracting the leptonic tensor 
in~(\ref{e:leptontensor}) with the hadronic tensor one finds the following
general form of the cross section in Eq.~(\ref{e:xs_2}):
\begin{align} \label{e:ang_dist}
& \frac{d\sigma}{d^4 q \, d \Omega} = \frac{\alpha_{em}^2}{F \, q^2} \times 
\nonumber \\
& \quad \Big \{ \Big(
   (1 + \cos^2 \theta) \, F_{UU}^{1} 
 + (1 - \cos^2 \theta) \, F_{UU}^{2} 
 + \sin 2\theta \cos \phi \, F_{UU}^{\cos \phi} 
 + \sin^2 \theta \cos 2\phi \, F_{UU}^{\cos 2\phi} \Big)
\nonumber \\
& \quad \; + \, S_{aL} \Big( 
   \sin 2\theta \sin \phi \, F_{LU}^{\sin \phi} 
 + \sin^2 \theta \sin 2\phi \, F_{LU}^{\sin 2\phi} \Big)
\nonumber \\
& \quad \; + \, S_{bL} \Big( 
   \sin 2\theta \sin \phi \, F_{UL}^{\sin \phi} 
 + \sin^2 \theta \sin 2\phi \, F_{UL}^{\sin 2\phi} \Big)
\nonumber \displaybreak[0] \\
& \quad \; + \, |\vec{S}_{aT}| \Big[ \sin \phi_{a} \Big( 
   (1 + \cos^2 \theta) \, F_{TU}^{1} 
 + (1 - \cos^2 \theta) \, F_{TU}^{2} 
 + \sin 2\theta \cos \phi \, F_{TU}^{\cos \phi} 
 + \sin^2 \theta \cos 2\phi \, F_{TU}^{\cos 2\phi} \Big)    
\nonumber \\
& \hspace{1.5cm} 
 + \cos \phi_{a} \Big(  
   \sin 2\theta \sin \phi \, F_{TU}^{\sin \phi} 
 + \sin^2 \theta \sin 2\phi \, F_{TU}^{\sin 2\phi} \Big) \Big]
\nonumber \displaybreak[0] \\
& \quad \; + \, |\vec{S}_{bT}| \Big[ \sin \phi_{b} \Big( 
   (1 + \cos^2 \theta) \, F_{UT}^{1} 
 + (1 - \cos^2 \theta) \, F_{UT}^{2} 
 + \sin 2\theta \cos \phi \, F_{UT}^{\cos \phi} 
 + \sin^2 \theta \cos 2\phi \, F_{UT}^{\cos 2\phi} \Big)    
\nonumber \\
& \hspace{1.5cm} 
 + \cos \phi_{b} \Big(  
   \sin 2\theta \sin \phi \, F_{UT}^{\sin \phi} 
 + \sin^2 \theta \sin 2\phi \, F_{UT}^{\sin 2\phi} \Big) \Big]
\nonumber \displaybreak[0] \\
& \quad \; + \, S_{aL} \, S_{bL} \Big(
   (1 + \cos^2 \theta) \, F_{LL}^{1} 
 + (1 - \cos^2 \theta) \, F_{LL}^{2} 
 + \sin 2\theta \cos \phi \, F_{LL}^{\cos \phi} 
 + \sin^2 \theta \cos 2\phi \, F_{LL}^{\cos 2\phi} \Big)
\nonumber \displaybreak[0] \\
& \quad \; + \, S_{aL} \, |\vec{S}_{bT}| \Big[ 
   \cos \phi_{b} \Big(
   (1 + \cos^2 \theta) \, F_{LT}^{1} 
 + (1 - \cos^2 \theta) \, F_{LT}^{2} 
 + \sin 2\theta \cos \phi \, F_{LT}^{\cos \phi} 
 + \sin^2 \theta \cos 2\phi \, F_{LT}^{\cos 2\phi} \Big)
\nonumber \\
& \hspace{2.1cm} 
 + \sin \phi_{b} \Big(  
   \sin 2\theta \sin \phi \, F_{LT}^{\sin \phi} 
 + \sin^2 \theta \sin 2\phi \, F_{LT}^{\sin 2\phi} \Big) \Big]
\nonumber \displaybreak[0] \\
& \quad \; + \, |\vec{S}_{aT}| \, S_{bL} \Big[ 
   \cos \phi_{a} \Big(
   (1 + \cos^2 \theta) \, F_{TL}^{1} 
 + (1 - \cos^2 \theta) \, F_{TL}^{2} 
 + \sin 2\theta \cos \phi \, F_{TL}^{\cos \phi} 
 + \sin^2 \theta \cos 2\phi \, F_{TL}^{\cos 2\phi} \Big)
\nonumber \\
& \hspace{2.1cm} 
 + \sin \phi_{a} \Big(  
   \sin 2\theta \sin \phi \, F_{TL}^{\sin \phi} 
 + \sin^2 \theta \sin 2\phi \, F_{TL}^{\sin 2\phi} \Big) \Big]
\nonumber \displaybreak[0] \\
& \quad \; + \, |\vec{S}_{aT}| \, |\vec{S}_{bT}|  \Big[ 
   \cos (\phi_{a} + \phi_{b}) \Big(
   (1 + \cos^2 \theta) \, F_{TT}^{1} 
 + (1 - \cos^2 \theta) \, F_{TT}^{2} 
 + \sin 2\theta \cos \phi \, F_{TT}^{\cos \phi}
 + \sin^2 \theta \cos 2\phi \, F_{TT}^{\cos 2\phi} \Big) 
\nonumber \\
& \hspace{2.3cm} 
 + \cos (\phi_a - \phi_b) \Big(  
   (1 + \cos^2 \theta) \, \bar{F}_{TT}^{1} 
 + (1 - \cos^2 \theta) \, \bar{F}_{TT}^{2}     
 + \sin 2\theta \cos \phi \, \bar{F}_{TT}^{\cos \phi}
 + \sin^2 \theta \cos 2\phi \, \bar{F}_{TT}^{\cos 2\phi} \Big) 
\nonumber \\
& \hspace{2.3cm} 
 + \sin (\phi_a + \phi_b) \Big(  
   \sin 2\theta \sin \phi \, F_{TT}^{\sin \phi} 
 + \sin^2 \theta \sin 2\phi \, F_{TT}^{\sin 2\phi} \Big)
\nonumber \\
& \hspace{2.3cm} 
 + \sin (\phi_a - \phi_b) \Big(  
   \sin 2\theta \sin \phi \, \bar{F}_{TT}^{\sin \phi} 
 + \sin^2 \theta \sin 2\phi \, \bar{F}_{TT}^{\sin 2\phi} \Big) \Big] \Big\} \,.
\end{align}
In Eq.~(\ref{e:ang_dist}) 48 structure functions show up which exactly matches with 
the number of the $V_i$ defined in Section III.
The structure functions again depend on the three variables  
$P_a \cdot q$, $P_b \cdot q$, and $q^2$, i.e.,
$F_{UU}^{1} = F_{UU}^{1}(P_a \cdot q ,\, P_b \cdot q ,\, q^2)$ and so on.
We refrain from giving the explicit relations between the structure functions 
in~(\ref{e:ang_dist}) and the $V_i$ because these lengthy formulae are not needed for 
the following discussion.
In order to shorten the notation in~(\ref{e:ang_dist}) we left out indices for the 
angles which characterize the lepton momenta and the transverse spin vectors of the 
hadrons.
There is yet another reason for omitting those indices:
the form of the angular distribution in~(\ref{e:ang_dist}) holds for any dilepton rest 
frame and not just the CS-frame.
The numerical values of the structure functions of course change when going from one 
frame to another.
Furthermore, note that the components of the spin vectors can be understood in different 
frames like the rest frame of one of the hadrons, the {\it cm}-frame, or a dilepton 
rest frame.

In particular for the angular distribution of the unpolarized cross section different
notations can be found in the literature (see, e.g.,~\cite{boer_06} and references 
therein).
Here we just quote the frequently used formula
\begin{equation}
\frac{dN}{d\Omega} \equiv 
\frac{d\sigma}{d^4 q \, d \Omega} \bigg / \frac{d\sigma}{d^4 q}
= \frac{3}{4\pi} \, \frac{1}{\lambda + 3} 
 \bigg( 1 + \lambda \cos^2 \theta 
      + \mu \sin 2\theta \cos \phi 
      + \frac{\nu}{2} \sin^2 \theta \cos 2\phi \bigg) \,.
\end{equation}
One readily finds
\begin{equation}
\lambda = \frac{F_{UU}^{1} - F_{UU}^{2}}{F_{UU}^{1} + F_{UU}^{2}} \,, \qquad
\mu = \frac{F_{UU}^{\cos \phi}}{F_{UU}^{1} + F_{UU}^{2}} \,, \qquad
\nu = \frac{2 \, F_{UU}^{\cos 2\phi}}{F_{UU}^{1} + F_{UU}^{2}} \,.
\end{equation}
The so-called Lam-Tung relation~\cite{lam_78,lam_80,collins_78}
\begin{equation} \label{e:lamtung}
\lambda + 2 \nu = 1 \,,
\end{equation}
which in terms of the structure functions defined in~(\ref{e:ang_dist}) reads
\begin{equation}
F_{UU}^{2} = 2 \, F_{UU}^{\cos 2\phi} \,,
\end{equation}
has attracted considerable attention in the past.
This relation is exact if one computes the DY process to ${\cal O}(\alpha_s)$ 
in the standard collinear perturbative QCD framework.
Even at ${\cal O}(\alpha_s^2)$ the numerical violation of~(\ref{e:lamtung}) is 
small~\cite{mirkes_94}.
On the other hand data for $\pi^- \, N \to \mu^- \, \mu^+ \, X$ taken at 
CERN~\cite{falciano_86,guanziroli_87} and at Fermilab~\cite{conway_89} are in 
disagreement with the Lam-Tung relation.
In particular, an unexpectedly large $\cos 2\phi$ modulation of the cross section was 
observed, and in the meantime different explanations for this phenomenon have been put 
forward in the 
literature~\cite{berger_79,brandenburg_93,brandenburg_94,boer_04,brandenburg_06,bakulev_07,hoyer_08}.
In Ref.~\cite{boer_99} it was pointed out that intrinsic transverse motion of initial 
state partons might be responsible for the observed violation of the Lam-Tung relation.
In the following section we will briefly return to this point in connection with the 
parton model calculation.
It is also worthwhile to mention that more recent Fermilab data on proton-deuteron 
Drell-Yan do agree with the Lam-Tung relation~\cite{zhu_06}.

The hadronic tensor given in Section III also allows one to find the angular 
distribution of the cross section for the specific kinematical point $q_T = 0$.
Altogether, in that case one has nine independent angular dependences and structure 
functions,
\begin{align} \label{e:ang_dist_0}
& \frac{d\sigma}{d^4 q \, d \Omega} \Big|_{q_T=0} = \frac{\alpha_{em}^2}{F \, q^2} \times 
\nonumber \\
& \quad \Big \{ \Big( 
   (1 + \cos^2 \theta) \, F_{UU}^{1} 
 + (1 - \cos^2 \theta) \, F_{UU}^{2} \Big) 
 + S_{aL} \, S_{bL} \Big(
   (1 + \cos^2 \theta) \, F_{LL}^{1} 
 + (1 - \cos^2 \theta) \, F_{LL}^{2} \Big)
\nonumber \\
& \quad \; + \, S_{aL} \, |\vec{S}_{bT}| \Big( 
   \sin 2\theta \cos (\phi - \phi_{b}) \, \tfrac{1}{2} \,
   (F_{LT}^{\cos \phi} + F_{LT}^{\sin \phi}) \Big)
 + |\vec{S}_{aT}| \, S_{bL} \Big( 
   \sin 2\theta \cos (\phi - \phi_{a}) \, \tfrac{1}{2} \,
   (F_{TL}^{\cos \phi} + F_{TL}^{\sin \phi}) \Big) 
\nonumber \\
& \quad \; + \, |\vec{S}_{aT}| \, |\vec{S}_{bT}|  \Big[ 
   \sin^2\theta \cos (2 \phi - \phi_{a} - \phi_{b}) \, \tfrac{1}{2} \,
   (F_{TT}^{\cos 2\phi} + F_{TT}^{\sin 2\phi})
\nonumber \\
& \hspace{2.3cm}
 + \cos (\phi_a - \phi_b) \Big(  
   (1 + \cos^2 \theta) \, \bar{F}_{TT}^{1} 
 + (1 - \cos^2 \theta) \, \bar{F}_{TT}^{2} \Big) \Big] \Big\} \,.
\end{align}
This result was already given in Ref.~\cite{ralston_79} using a different notation. 
We note that in the double-polarized sector the following relations hold:
\begin{equation}
F_{LT}^{\cos \phi} \big|_{q_T = 0} = F_{LT}^{\sin \phi} \big|_{q_T = 0} \,, 
\qquad 
F_{TL}^{\cos \phi} \big|_{q_T = 0} = F_{TL}^{\sin \phi} \big|_{q_T = 0} \,,
\qquad
F_{TT}^{\cos 2\phi} \big|_{q_T = 0} = F_{TT}^{\sin 2\phi} \big|_{q_T = 0} \,.
\end{equation}
All the structure functions that show up in~(\ref{e:ang_dist}) but not 
in~(\ref{e:ang_dist_0}) have a kinematical zero at $q_T = 0$.
Notice that the angular distribution of the $q_T$-integrated cross section agrees
with the one in~(\ref{e:ang_dist_0}), but the corresponding structure functions 
differ numerically (see also~\cite{ralston_79}).

We have pointed out that our hadronic tensor provides the angular distribution
of the cross section at $q_T = 0$.
In fact, this statement is not totally correct.
The hadronic tensor in Eq.~(\ref{e:tensorba_ab_1}) does not generate the term 
proportional to $\cos (2 \phi - \phi_a - \phi_b)$ in~(\ref{e:ang_dist_0}).
This can be understood by taking a close look at Eq.~(\ref{e:detrel_b}): there 
the prefactor in front of $h_{ab,48}^{\mu\nu}$ vanishes for $q_T = 0$ implying
that at this particular kinematical point one is actually not allowed to eliminate 
this tensor which does generate the required $\cos (2 \phi - \phi_a - \phi_b)$ 
term for $q_T = 0$.
One may instead eliminate, for instance, the tensor $h_{ab,47}^{\mu\nu}$.
Since the rather specific case $q_T = 0$ was already worked out in the 
literature~\cite{ralston_79} we have proposed the tensor in~(\ref{e:tensorba_ab_1}) 
for the sake of symmetry.

In Section III.D we have considered the case of identical hadrons in the initial 
state and the resultant constraints for the structure functions $V_i$.
One can do a corresponding analysis for the structure functions defined in 
Eq.~(\ref{e:ang_dist}).
The key ingredient of such an analysis is that the cross section remains the same 
if the hadrons are exchanged. 
Note that the exchange $H_a \leftrightarrow H_b$ also leads to the reversal of the
$z$-direction which, in particular, implies
\begin{equation}
\phi_a \leftrightarrow - \phi_b \,, \qquad
\phi \to - \phi \,, \qquad
\theta \to \pi - \theta \,.
\end{equation}
Twenty structure functions are either symmetric or antisymmetric under the exchange 
$P_a \leftrightarrow P_b$.
Using the shorthand notation of Eqs.~(\ref{e:sym_v}),~(\ref{e:rel_v}) one finds
\begin{equation} \label{e:sym_f}
\begin{array}{ll}
F_{UU}^{1}(b,a) = F_{UU}^{1}(a,b) \,, \qquad &
F_{UU}^{2}(b,a) = F_{UU}^{2}(a,b) \,, 
\\[0.2cm]
F_{UU}^{\cos \phi}(b,a) = - F_{UU}^{\cos \phi}(a,b) \,, \qquad &
F_{UU}^{\cos 2\phi}(b,a) = F_{UU}^{\cos 2\phi}(a,b) \,,
\\[0.2cm]
F_{LL}^{1}(b,a) = F_{LL}^{1}(a,b) \,, \qquad &
F_{LL}^{2}(b,a) = F_{LL}^{2}(a,b) \,, 
\\[0.2cm]
F_{LL}^{\cos \phi}(b,a) = - F_{LL}^{\cos \phi}(a,b) \,, \qquad &
F_{LL}^{\cos 2\phi}(b,a) = F_{LL}^{\cos 2\phi}(a,b) \,,
\\[0.2cm]
F_{TT}^{1}(b,a) = F_{TT}^{1}(a,b) \,, \qquad &
F_{TT}^{2}(b,a) = F_{TT}^{2}(a,b) \,, 
\\[0.2cm]
F_{TT}^{\cos \phi}(b,a) = - F_{TT}^{\cos \phi}(a,b) \,, \qquad &
F_{TT}^{\cos 2\phi}(b,a) = F_{TT}^{\cos 2\phi}(a,b) \,,
\\[0.2cm]
\bar{F}_{TT}^{1}(b,a) = \bar{F}_{TT}^{1}(a,b) \,, \qquad &
\bar{F}_{TT}^{2}(b,a) = \bar{F}_{TT}^{2}(a,b) \,, 
\\[0.2cm]
\bar{F}_{TT}^{\cos \phi}(b,a) = - \bar{F}_{TT}^{\cos \phi}(a,b) \,, \qquad &
\bar{F}_{TT}^{\cos 2\phi}(b,a) = \bar{F}_{TT}^{\cos 2\phi}(a,b) \,,
\\[0.2cm]
F_{TT}^{\sin \phi}(b,a) = - F_{TT}^{\sin \phi}(a,b) \,, \qquad &
F_{TT}^{\sin 2\phi}(b,a) = F_{TT}^{\sin 2\phi}(a,b) \,,
\\[0.2cm]
\bar{F}_{TT}^{\sin \phi}(b,a) = \bar{F}_{TT}^{\sin \phi}(a,b) \,, \qquad &
\bar{F}_{TT}^{\sin 2\phi}(b,a) = - \bar{F}_{TT}^{\sin 2\phi}(a,b) \,.
\end{array}
\end{equation}
The remaining structure functions fulfil the relations
\begin{equation} \label{e:rel_f}
\begin{array}{ll}
F_{UL}^{\sin \phi}(b,a) = F_{LU}^{\sin \phi}(a,b) \,, \qquad &
F_{UL}^{\sin 2\phi}(b,a) = - F_{LU}^{\sin 2\phi}(a,b) \,, 
\\[0.2cm]
F_{UT}^{1}(b,a) = - F_{TU}^{1}(a,b) \,, \qquad &
F_{UT}^{2}(b,a) = - F_{TU}^{2}(a,b) \,,
\\[0.2cm]
F_{UT}^{\cos \phi}(b,a) = F_{TU}^{\cos \phi}(a,b) \,, \qquad &
F_{UT}^{\cos 2\phi}(b,a) = - F_{TU}^{\cos 2\phi}(a,b) \,,
\\[0.2cm]
F_{UT}^{\sin \phi}(b,a) = F_{TU}^{\sin \phi}(a,b) \,, \qquad &
F_{UT}^{\sin 2\phi}(b,a) = - F_{TU}^{\sin 2\phi}(a,b) \,,
\\[0.2cm]
F_{TL}^{1}(b,a) = F_{LT}^{1}(a,b) \,, \qquad &
F_{TL}^{2}(b,a) = F_{LT}^{2}(a,b) \,, 
\\[0.2cm]
F_{TL}^{\cos \phi}(b,a) = - F_{LT}^{\cos \phi}(a,b) \,, \qquad &
F_{TL}^{\cos 2\phi}(b,a) = F_{LT}^{\cos 2\phi}(a,b) \,,
\\[0.2cm]
F_{TL}^{\sin \phi}(b,a) = - F_{LT}^{\sin \phi}(a,b) \,, \qquad &
F_{TL}^{\sin 2\phi}(b,a) = F_{LT}^{\sin 2\phi}(a,b) \,.
\end{array}
\end{equation}
It is of course intuitively clear that for identical hadrons relations as given
in~(\ref{e:rel_f}) have to exist.
But one has to keep in mind that relative signs between the corresponding
structure functions can show up.
Eventually, we mention that~(\ref{e:sym_f}),~(\ref{e:rel_f}) can also be derived 
from~(\ref{e:sym_v}),~(\ref{e:rel_v}) and the relations between the two sets of
structure functions.

\section{Parton model approximation}

This section deals with the parton model description of the structure functions in 
Eq.~(\ref{e:ang_dist}). 
Up to this point we didnot specify the external kinematics of the process.
In the following we will consider the kinematical regime where the transverse 
photon momentum $q_T$ is of the order of a typical hadronic mass scale which means, 
in particular, that it is much smaller than the hard scale $q$.
This is the region where TMDs enter the description of the DY process in a natural 
way.

Our treatment is restricted to leading twist, i.e., to the leading order of an 
expansion in powers of $1/q$.
Mainly because of the potential problems of subleading twist TMD-factorization 
pointed out in Refs.~\cite{gamberg_06,bacchetta_08} we refrain here from including 
the twist-3 case.
Moreover, we neither take into account higher order hard scattering corrections
nor effects associated with soft gluon radiation.
For three of the structure functions such contributions were considered 
in~\cite{ji_04b}.

\subsection{Hadronic tensor}

The parton model description of the Drell-Yan process can be represented by the 
diagrams shown in Fig.~\ref{f:dy_amp}, where, e.g., the scattering amplitude for 
diagram (a) reads
\begin{eqnarray} \label{e:dy_amp}
i M_{(a)} = \sum_{q} \sum_{c=1}^{N_{c}} \, \frac{i e_{q} e^{2}}{q^{2}} \, 
\langle X_{a} | \, \psi_{i}^{c,q}(0) \, |P_{a}, S_{a} \rangle \, 
\langle X_{b} | \, \bar{\psi}_{j}^{c,q}(0)\,|P_{b}, S_{b} \rangle
\Big[ (\gamma^{\mu})_{ji} \,
      \bar{u}(l,\lambda) \, \gamma_{\mu} \, v(l^{\prime},\lambda^{\prime}) \Big] \,.
\end{eqnarray}
A sum over color $c$ and the quark flavors $q$ is implemented explicitly in this 
expression. 
The electromagnetic charge of the quark, in units of the elementary charge $e$,
is denoted by $e_{q}$.
A corresponding formula holds for the amplitude $M_{(b)}$ of the graph (b).
The differential cross section~(\ref{e:xs_2}) in a dilepton rest frame is then 
given by
\begin{equation}
\frac{d\sigma}{d^4 q \, d\Omega} =
\frac{1}{8 \, (2\pi)^{2} \, F} \, \sum_{\lambda,\lambda^{\prime}} \, 
\sum_{X_{a},X_{b}} \hspace{-0.6cm} \int \hspace{0.2cm} 
\Big( \big| M_{(a)} \big|^{2} + \big| M_{(b)} \big|^2 \Big) \,
\delta^{(4)}(P_{X_{a}} + P_{X_{b}} + q - P_a - P_b) \,.
\end{equation}
Note that there is no interference between the two diagrams in Fig.~\ref{f:dy_amp}. 
One can modify this formula by introducing the momenta of the active partons, 
$k_{a}$ and $k_{b}$. 
This allows one to sum over a complete set of intermediate states and to rewrite 
the hadronic part of the cross section in terms of fully unintegrated quark-quark 
correlators (see, e.g., 
Ref.~\cite{tangerman_94a,mulders_95,boer_97c,goeke_05,bacchetta_06,collins_07,meissner_08a}).
In doing so one finds the hadronic tensor
\begin{equation} \label{e:ht_pm}
W^{\mu \nu} = \frac{1}{N_c} \sum_q e_q^2 \int d^4 k_a \, d^4 k_b \, 
\delta^{(4)}(q - k_a - k_b) \, 
\mathrm{Tr} \big[ \gamma^{\mu} \, \Phi^q(k_a, P_a, S_a | n_a) \, 
     \gamma^{\nu} \, \bar{\Phi}^q(k_b, P_b, S_b | n_b) \big]
+\{\Phi \leftrightarrow \bar{\Phi}\} \,,
\end{equation}
where the quark-quark correlators, which depend on the full 4-momentum of the quarks,
are defined as
\begin{eqnarray} \label{e:fullphi_a}
\Phi_{ij}^{q}(k_{a}, P_{a}, S_{a} | n_{a}) & = & 
\int \frac{d^{4} z}{(2\pi)^{4}} \, e^{ik_{a} \cdot z} \, 
\langle P_{a}, S_{a} | \, \bar{\psi}_{j}^{q}(0) \, \mathcal{W}_{\textrm{DY}}[0,z|n_{a}] \, 
        \psi_{i}^{q}(z) \, | P_{a}, S_{a} \rangle \,, 
\\ \label{e:fullphi_b}
\bar{\Phi}_{ij}^{q}(k_{b}, P_{b}, S_{b} | n_{b}) & = & 
\int \frac{d^{4} z}{(2\pi)^{4}} \, e^{ik_{b} \cdot z} \, 
\langle P_{b}, S_{b} | \, \psi_{i}^{q}(0) \, \mathcal{W}_{\textrm{DY}}[0,z|n_{b}] \, 
        \bar{\psi}_{j}^{q}(z) \, | P_{b}, S_{b} \rangle \,.
\end{eqnarray}
The object $\mathcal{W}$ denotes a gauge link operator (Wilson line) which ensures 
color gauge invariance of the correlators. 
We note that actually the Wilson lines cannot be derived from the diagrams in 
Fig.~\ref{f:dy_amp}.
They are generated, however, if in addition collinear gluon exchanges between the 
active partons and the remnants of the incoming hadrons are taken into account
(see, e.g., Refs.~\cite{collins_02,ji_02,belitsky_02,boer_03a}).
In general, the Wilson lines entering unintegrated parton correlators are 
process-dependent.
For the DY process we will specify them below but already emphasize here their 
dependence on a light-cone vector $n_{a}$ or $n_{b}$. 
Note that in Eqs.~(\ref{e:fullphi_a}),~(\ref{e:fullphi_b}) a color sum is implicit, 
leading to the factor $1/N_{c}=1/3$ in~(\ref{e:ht_pm}). 
The term $\{\Phi\leftrightarrow\bar{\Phi}\}$ in Eq.~(\ref{e:ht_pm}) represents the 
contribution of the diagram in Fig.~\ref{f:dy_amp} (b) and is obtained from the first 
term by interchanging the correlators. 

In the parton model initial state partons are assumed to move quasi-collinearly with 
respect to their parent hadron. 
Consequently, the components of the parton momenta behave like the corresponding 
components of the hadron momenta. 
The following estimates for the parton momenta in the DY process are valid in frames 
where the hadron $H_a$ has a large light-cone plus-momentum and the hadron $H_b$ a 
large minus-momentum (this applies, in particular, to the {\it cm}-frame --- 
see also the discussion in Section IV):
\begin{align} \label{e:mom_appr}
& k_{a}^{+} \sim {\cal O}(q) \,, \quad 
  k_{a}^{-} \sim {\cal O}(1/q) \,, \quad
  k_{aT} \sim {\cal O}(q^{0}) \,,
\nonumber \\
& k_{b}^{+} \sim {\cal O}(1/q) \,, \quad 
  k_{b}^{-} \sim {\cal O}(q) \,, \quad
  k_{bT} \sim {\cal O}(q^{0}) \,,
\end{align}
where we use the light-cone components $v^{\pm} = (v^0 \pm v^3)/\sqrt{2}$ for a 
generic 4-vector $v$.
From the standpoint of factorization this means that $\Phi$ and $\bar{\Phi}$ are 
treated as nonperturbative objects because the kinematical invariants
$k_{a} \cdot P_{a}, \; k_{a}^{2}, \; k_{b} \cdot P_{b}, \; k_{b}^{2}$ on which
the correlators depend are much smaller than $q^2$.
According to~(\ref{e:mom_appr}) the momentum components $k_{a}^{-}$ and $k_{b}^{+}$ are 
small and hence can be neglected in the $\delta$-function in Eq.~(\ref{e:ht_pm}).
This also automatically implies $q^+ \approx k_a^+$ and $q^- \approx k_b^-$.
The hadronic tensor then reduces to
\begin{align} \label{e:ht_pm_kt}
& W^{\mu \nu} = \frac{1}{N_c} \sum_q e_q^2 \int d^2 \vec{k}_{aT} \, d^2 \vec{k}_{bT} \, 
 \delta^{(2)}(\vec{q}_T - \vec{k}_{aT} - \vec{k}_{bT}) \, 
\mathrm{Tr} \big[ \gamma^{\mu} \, \Phi^q(x_a, \vec{k}_{aT}, S_a | n_a) \, 
     \gamma^{\nu} \, \bar{\Phi}^q(x_b, \vec{k}_{bT}, S_b | n_b) \big]
\nonumber \\
& \hspace{1.2cm} +\{\Phi \leftrightarrow \bar{\Phi}\} \,,
\end{align}
where we used the common DY variables 
\begin{equation}
x_a = \frac{q^2}{2P_a \cdot q} \approx \frac{k_a^+}{P_a^+} \,, \qquad 
x_b = \frac{q^2}{2P_b \cdot q} \approx \frac{k_b^-}{P_b^-} \,.
\end{equation}
The transverse momentum dependent quark-quark correlators in~(\ref{e:ht_pm_kt}) are 
defined according to
\begin{eqnarray} \label{e:phikt_a}
\Phi_{ij}^{q}(x_{a}, \vec{k}_{aT}, S_{a} | n_{a}) & = & 
\int \frac{dz^- \, d^2\vec{z}_T}{(2\pi)^{3}} \, e^{ik_{a} \cdot z} \, 
\langle P_{a}, S_{a} | \, \bar{\psi}_{j}^{q}(0) \, \mathcal{W}_{\textrm{DY}}[0,z|n_{a}] \, 
        \psi_{i}^{q}(z) \, | P_{a}, S_{a} \rangle \big|_{z^+ = 0} \,, 
\\ \label{e:phikt_b}
\bar{\Phi}_{ij}^{q}(x_{b}, \vec{k}_{bT}, S_{b} | n_{b}) & = & 
\int \frac{dz^+ \, d^2\vec{z}_T}{(2\pi)^{3}} \, e^{ik_{b} \cdot z} \, 
\langle P_{b}, S_{b} | \, \psi_{i}^{q}(0) \, \mathcal{W}_{\textrm{DY}}[0,z|n_{b}] \, 
        \bar{\psi}_{j}^{q}(z) \, | P_{b}, S_{b} \rangle \big|_{z^- =  0} \,,
\end{eqnarray}
and they are obtained from the correlators in~(\ref{e:fullphi_a}),~(\ref{e:fullphi_b}) 
by integrating out the respective small light-cone momentum of the parton.
We now specify the Wilson lines in the quark-quark correlators.
The appropriate choice for the DY process is~\cite{collins_02,ji_02,belitsky_02}
\begin{eqnarray} \label{e:wilson_a}
\mathcal{W}_{\mathrm{DY}}[0,z|n_{a}] \big|_{z^+ = 0} & = & 
[0 \, ; \, -\infty \, n_{a}] \times 
[-\infty \, n_{a} \, ; \, -\infty \, n_{a} + z_T] \times 
[-\infty \, n_{a} + z_T \, ; \, z^- n_{a} + z_T],
\vphantom{\frac{1}{1}} \\ \label{e:wilson_b}
\mathcal{W}_{\mathrm{DY}}[0,z|n_{b}] \big|_{z^- = 0} & = & 
[0 \, ; \, -\infty \, n_{b}] \times 
[-\infty \, n_{b} \, ; \, -\infty \, n_{b} + z_T] \times 
[-\infty \, n_{b} + z_T \, ; \, z^- n_{b} + z_T],
\end{eqnarray}
with $[a;b]$ denoting a straight gauge link between the positions $a$ and $b$, 
and $z_T^{\mu} \equiv (0,\vec{z}_T,0)$.
The light-cone vectors in~(\ref{e:wilson_a}),~(\ref{e:wilson_b}) are given by 
\begin{equation}
n_{a}^{\mu} = \frac{1}{\sqrt{2}} \, (1,0,0,-1) \,, \qquad 
n_{b}^{\mu} = \frac{1}{\sqrt{2}} \, (1,0,0,1) \,.
\end{equation}
Note that the diagram in Fig.~\ref{f:dy_amp}(b) generates, e.g., the correlator 
$\Phi^{q}(k_{b}, P_{b}, S_{b} | n_{b})$ which can be related to
$\Phi^{q}(k_{a}, P_{a}, S_{a} | n_{a})$ in Eq.~(\ref{e:fullphi_a}) by means
of the parity transformation.

We also mention that so-called light-cone divergences, which are caused by the 
light-like Wilson lines in~(\ref{e:wilson_a}),~(\ref{e:wilson_b}), can be avoided 
if near-light-cone directions for the Wilson lines are chosen instead.
For a discussion of such divergences and other nontrivial issues concerning the
precise definition of unintegrated parton correlation functions we refer to the
recent contribution~\cite{collins_08} as well as references therein.

\subsection{Transverse momentum dependent parton distributions}

The quark-quark correlators in Eqs.~(\ref{e:phikt_a}),~(\ref{e:phikt_b}) can be 
parameterized through
TMDs~\cite{ralston_79,tangerman_94a,mulders_95,boer_97c,goeke_05,bacchetta_06}.
A common and rather convenient procedure for performing such a parameterization 
is by specifying the traces of the correlators with the Dirac-matrices
$\Gamma = \gamma^{\mu}, \, \gamma^{\mu}\gamma_{5}, \, i\sigma^{\mu\nu}\gamma_{5},
\, 1, \, i\gamma_{5}$,
\begin{equation}
\Phi^{q \, [\Gamma]} \equiv \frac{1}{2} \, \mathrm{Tr} \, [ \Phi^{q} \, \Gamma] \,.
\end{equation}
In the {\it cm}-frame, where the hadron $H_a$ has a large plus-momentum, the 
leading (twist) traces are $\Phi^{[\gamma^{+}]}$, $\Phi^{[\gamma^{+}\gamma_{5}]}$,
and $\Phi^{[i\sigma^{i+}\gamma_{5}]}$ ($i=\{1,2\}$), while all the other traces
are suppressed in the cross section by at least one power of the large light-cone 
momentum (and consequently by one power of $q$).
These traces then have the following expressions in terms of leading twist quark
TMDs (see, e.g.,~\cite{goeke_05,bacchetta_06}\footnote{Note that the {\it l.h.s.}
in Eq.~(16) of~\cite{goeke_05} should read $\Phi^{[i\sigma^{i+}\gamma_5]}$.}):
\begin{align} \label{e:phiq_1}
\Phi^{q \, [\gamma^{+}]} & = \, 
 f_{1}^{q}(x_{a},\vec{k}_{aT}^{2}) - 
 \frac{\varepsilon_{T}^{ij} k_{aT}^{i} S_{aT}^{j}}{M_{a}} \, 
 f_{1T}^{\perp q}(x_{a},\vec{k}_{aT}^{2}) \,, 
\displaybreak[0] \\ \label{e:phiq_2}
\Phi^{q \, [\gamma^{+}\gamma_{5}]} & = \, 
 S_{aL} \, g_{1L}^{q}(x_{a},\vec{k}_{aT}^{2}) +
 \frac{\vec{k}_{aT} \cdot \vec{S}_{aT}}{M_{a}} \, 
 g_{1T}^{q}(x_{a},\vec{k}_{aT}^{2}) \,,
\displaybreak[0] \\ \label{e:phiq_3}
\Phi^{q \, [i\sigma^{i+}\gamma_{5}]} & = \, 
 S_{aT}^{i} \, h_{1}^{q}(x_{a},\vec{k}_{aT}^{2}) + 
 \frac{k_{aT}^{i}(\vec{k}_{aT}\cdot\vec{S}_{aT}) - 
       \frac{1}{2}\vec{k}_{aT}^{2}S_{aT}^{i}}{M_{a}^{2}} \, 
 h_{1T}^{\perp q}(x_{a},\vec{k}_{aT}^{2})
\nonumber \\
 & \hspace{0.5cm} 
 + S_{aL} \, \frac{k_{aT}^{i}}{M_{a}} \, h_{1L}^{\perp q}(x_{a},\vec{k}_{aT}^{2}) +
 \frac{\varepsilon_{T}^{ij} k_{aT}^{j}}{M_{a}} \, h_{1}^{\perp q}(x_{a},\vec{k}_{aT}^{2}) \,.
\end{align}
For brevity we omitted the arguments of the correlator $\Phi$.
Note that the components of the nucleon spin vector in (\ref{e:phiq_1})--(\ref{e:phiq_3}) 
are understood in the {\it cm}-frame.
The object $\varepsilon_{T}^{ij}$ represents a short form of the transverse 
epsilon tensor $\varepsilon^{-+ij}$, where we use the convention 
$\varepsilon^{-+12} = 1$. 
The transverse momentum dependent unpolarized quark distribution, helicity distribution,
and transversity distribution are denoted by $f_{1}$, $g_{1L}$, and $h_{1}$,
respectively.
Of particular importance are also the time-reversal odd (T-odd) Sivers function
$f_{1T}^{\perp}$~\cite{sivers_89,sivers_90} and Boer-Mulders function 
$h_{1}^{\perp}$~\cite{boer_97c} as they can give rise to quite interesting single spin 
and/or azimuthal asymmetries in hard semi-inclusive reactions.

The correlator $\bar{\Phi}^{q}$ in Eq.~(\ref{e:phikt_b}) is related to the correlator 
$\Phi^{\bar{q}}$ which defines, precisely in analogy to the 
Eqs.~(\ref{e:phiq_1})--(\ref{e:phiq_3}), antiquark distributions. 
For the different Dirac traces the relation reads~\cite{tangerman_94a}
\begin{equation} \label{e:rel_anti}
\bar{\Phi}^{q \, [\Gamma]} = \pm \, \Phi^{\bar{q} \, [\Gamma]} \,, \quad
\left\{ 
\begin{array}{l}
+ \; \mathrm{for} \; \gamma^{\mu}, \; i\sigma^{\mu\nu}\gamma_{5} \\
- \; \mathrm{for} \; \gamma^{\mu}\gamma_{5}, \; 1, \; i\gamma_{5} 
\end{array} \right. .
\end{equation}
Since the correlator $\bar{\Phi}$ in~(\ref{e:phikt_b}) is associated with the hadron
$H_b$ having a large minus-momentum in the \emph{cm}-frame, the leading traces are 
now $\bar{\Phi}^{[\gamma^{-}]}$, $\bar{\Phi}^{[\gamma^{-}\gamma_{5}]}$,
and $\bar{\Phi}^{[i\sigma^{i-}\gamma_{5}]}$. 
Taking~(\ref{e:rel_anti}) into account the parameterizations can be directly obtained
from (\ref{e:phiq_1})--(\ref{e:phiq_3}),
\begin{align} \label{e:phia_1}
\bar{\Phi}^{q \, [\gamma^{-}]} & = \, 
 f_{1}^{\bar{q}}(x_{b},\vec{k}_{bT}^{2}) + 
 \frac{\varepsilon_{T}^{ij} k_{bT}^{i} S_{bT}^{j}}{M_{b}} \, 
 f_{1T}^{\perp \bar{q}}(x_{b},\vec{k}_{bT}^{2}) \,, 
\displaybreak[0] \\ \label{e:phia_2}
\bar{\Phi}^{q \, [\gamma^{-}\gamma_{5}]} & = \, 
 - S_{bL} \, g_{1L}^{\bar{q}}(x_{b},\vec{k}_{bT}^{2}) -
 \frac{\vec{k}_{bT} \cdot \vec{S}_{bT}}{M_{b}} \, 
 g_{1T}^{\bar{q}}(x_{b},\vec{k}_{bT}^{2}) \,,
\displaybreak[0] \\ \label{e:phia_3}
\bar{\Phi}^{q \, [i\sigma^{i-}\gamma_{5}]} & = \, 
 S_{bT}^{i} \, h_{1}^{\bar{q}}(x_{b},\vec{k}_{bT}^{2}) + 
 \frac{k_{bT}^{i}(\vec{k}_{bT}\cdot\vec{S}_{bT}) - 
       \frac{1}{2}\vec{k}_{bT}^{2}S_{bT}^{i}}{M_{b}^{2}} \, 
 h_{1T}^{\perp \bar{q}}(x_{b},\vec{k}_{bT}^{2})
\nonumber \\
 & \hspace{0.5cm} 
 + S_{bL} \, \frac{k_{bT}^{i}}{M_{b}} \, h_{1L}^{\perp \bar{q}}(x_{b},\vec{k}_{bT}^{2}) -
 \frac{\varepsilon_{T}^{ij} k_{bT}^{j}}{M_{b}} \, h_{1}^{\perp \bar{q}}(x_{b},\vec{k}_{bT}^{2}) \,.
\end{align}
Note the respective sign change in front of the epsilon tensor $\varepsilon_{T}^{ij}$ which 
is due to the interchange of plus-momenta and minus-momenta.

\subsection{Leading spin observables}

Now we are in a position to calculate all the leading twist observables for $q_T \ll q$ 
by inserting the traces (\ref{e:phiq_1})--(\ref{e:phiq_3}) and 
(\ref{e:phia_1})--(\ref{e:phia_3}) into the hadronic tensor (\ref{e:ht_pm_kt}).
We mention again that the contraction of the hadronic and the leptonic tensor is 
performed in the {\it cm}-frame where, in order to get the leptonic tensor in that 
frame, use is made of Eqs.~(\ref{e:lepton1_cm}),~(\ref{e:lepton2_cm}).
Since the lepton momenta contain angles in the CS-frame our final result for the 
cross section is of the form~(\ref{e:ang_dist}).
Here one has to keep in mind that the leading twist calculation of course merely 
provides nonzero results for part of the structure functions in (\ref{e:ang_dist}).
Carrying out the contraction of the tensors and keeping only the leading contribution 
in $1/q$ one finds
\begin{eqnarray}
\frac{d\sigma}{d^4 q \, d\Omega} & = & 
\frac{\alpha_{em}^2 \, x_{a}x_{b}}{2 \, q^{4}} \, \frac{1}{N_c} \, \sum_{q} e_{q}^{2} 
\int d^{2}\vec{k}_{aT} \, d^{2}\vec{k}_{bT} \, 
\delta^{(2)}(\vec{q}_T - \vec{k}_{aT} - \vec{k}_{bT}) \times
\nonumber \\
&  & \Big[ (1+\cos^{2}\theta) \Big( \Phi^{q \, [\gamma^{+}]} \, \bar{\Phi}^{q \, [\gamma^{-}]}
 + \Phi^{q \, [\gamma^{+}\gamma_{5}]} \, \bar{\Phi}^{q \, [\gamma^{-}\gamma_{5}]} \Big)
\nonumber \\
&  & + \sin^{2}\theta \, 
 \Big( \cos 2\phi \, \big( \delta^{i1} \delta^{j1} - \delta^{i2} \delta^{j2} \big)
     + \sin 2\phi \, \big( \delta^{i1} \delta^{j2} + \delta^{i2} \delta^{j1} \big) \Big) \, 
 \Phi^{q \, [i\sigma^{i+}\gamma_{5}]} \, \bar{\Phi}^{q \, [i\sigma^{j-}\gamma_{5}]} \Big]
\nonumber \\
&  & +\{\Phi \leftrightarrow \bar{\Phi}\} + \mathcal{O}(1/q) \,.
\end{eqnarray}
To present the leading twist spin observables we will make use of the following notation 
for the convolution of TMDs in the transverse momentum space:
\begin{eqnarray} \label{e:convol}
{\cal C} \, [w(\vec{k}_{aT},\vec{k}_{bT})f_{1}\bar{f}_{2}] & \equiv & 
\frac{1}{N_c} \, \sum_{q} \, e_{q}^{2} \, 
\int d^{2}\vec{k}_{aT} \, d^{2}\vec{k}_{bT} \,
\delta^{(2)}(\vec{q}_T - \vec{k}_{aT} - \vec{k}_{bT}) \,
w(\vec{k}_{aT},\vec{k}_{bT}) \times 
\nonumber \\
& & \hspace{2cm}
\Big[f_{1}^{q}(x_{a},\vec{k}_{aT}^{2}) \, f_{2}^{\bar{q}}(x_{b},\vec{k}_{bT}^{2}) +
     f_{1}^{\bar{q}}(x_{a},\vec{k}_{aT}^{2}) \, f_{2}^{q}(x_{b},\vec{k}_{bT}^{2}) \Big] \,.
\end{eqnarray}
The two terms on the {\it r.h.s.} of~(\ref{e:convol}) are generated by the two diagram in
Fig.~\ref{f:dy_amp}.
For the parton model calculation it is convenient to introduce a number of linear 
combinations of various structure functions given in Eq.~(\ref{e:ang_dist}):
\begin{equation}
\begin{array}{ll}
F_{TU}^{\sin (2\phi - \phi_a)} \, \equiv \,
- \frac{1}{2} \Big( F_{TU}^{\cos 2\phi} - F_{TU}^{\sin 2\phi} \Big) \,, 
\qquad &
F_{TU}^{\sin (2\phi + \phi_a)} \, \equiv \,
\frac{1}{2} \Big( F_{TU}^{\cos 2\phi} + F_{TU}^{\sin 2\phi} \Big) \,,
\\[0.3cm]
F_{UT}^{\sin (2\phi - \phi_b)} \, \equiv \,
- \frac{1}{2} \Big( F_{UT}^{\cos 2\phi} - F_{UT}^{\sin 2\phi} \Big) \,, 
\qquad &
F_{UT}^{\sin (2\phi + \phi_b)} \, \equiv \,
\frac{1}{2} \Big( F_{UT}^{\cos 2\phi} + F_{UT}^{\sin 2\phi} \Big) \,,
\\[0.3cm]
F_{LT}^{\cos (2\phi - \phi_b)} \, \equiv \, 
\frac{1}{2} \Big( F_{LT}^{\cos 2\phi} + F_{LT}^{\sin 2\phi} \Big) \,, 
\qquad &
F_{LT}^{\cos (2\phi + \phi_b)} \, \equiv \,
\frac{1}{2} \Big( F_{LT}^{\cos 2\phi} - F_{LT}^{\sin 2\phi} \Big) \,,
\\[0.3cm]
F_{TL}^{\cos (2\phi - \phi_a)} \, \equiv \, 
\frac{1}{2} \Big( F_{TL}^{\cos 2\phi} + F_{TL}^{\sin 2\phi} \Big) \,, 
\qquad &
F_{TL}^{\cos (2\phi + \phi_a)} \, \equiv \,
\frac{1}{2} \Big( F_{TL}^{\cos 2\phi} - F_{TL}^{\sin 2\phi} \Big) \,,
\\[0.3cm]
F_{TT}^{\cos (2\phi - \phi_a - \phi_b)} \, \equiv \,
\frac{1}{2} \Big( F_{TT}^{\cos 2\phi} + F_{TT}^{\sin 2\phi} \Big) \,, 
\qquad &
F_{TT}^{\cos (2\phi - \phi_a + \phi_b)} \, \equiv \,
\frac{1}{2} \Big( \bar{F}_{TT}^{\cos 2\phi} + \bar{F}_{TT}^{\sin 2\phi} \Big) \,,
\\[0.3cm]
F_{TT}^{\cos (2\phi + \phi_a - \phi_b)} \, \equiv \,
\frac{1}{2} \Big( \bar{F}_{TT}^{\cos 2\phi} - \bar{F}_{TT}^{\sin 2\phi} \Big) \,, 
\qquad &
F_{TT}^{\cos (2\phi + \phi_a + \phi_b)} \, \equiv \,
\frac{1}{2} \Big( F_{TT}^{\cos 2\phi} - F_{TT}^{\sin 2\phi} \Big) \,.
\end{array}
\end{equation}
Using the unit vector $\vec{h} \equiv \vec{q}_T / q_T$ one eventually finds the 
following leading order structure functions in the CS-frame:
\begin{align} \label{e:uu_1}
F_{UU}^{1} & \, = \, 
{\cal C} \, \big[ f_{1} \, \bar{f}_{1} \big] \,, 
\vphantom{\Bigg[} \\ \label{e:uu_2}
F_{UU}^{\cos 2\phi} & \, = \, 
{\cal C} \, \Bigg[ \frac{2\big( \vec{h} \cdot \vec{k}_{aT} \big) 
                          \big( \vec{h} \cdot \vec{k}_{bT} \big)
                         - \vec{k}_{aT} \cdot \vec{k}_{bT}} {M_{a} M_{b}} \, 
                  h_{1}^{\perp} \, \bar{h}_{1}^{\perp} \Bigg] \,,
\displaybreak[0] \\
F_{LU}^{\sin 2\phi} & \, = \, 
{\cal C} \, \Bigg[ \frac{2\big( \vec{h} \cdot \vec{k}_{aT} \big) 
                          \big( \vec{h} \cdot \vec{k}_{bT} \big)
                         - \vec{k}_{aT} \cdot \vec{k}_{bT}} {M_{a} M_{b}} \, 
                  h_{1L}^{\perp} \, \bar{h}_{1}^{\perp} \Bigg] \,,
\displaybreak[0] \\
F_{UL}^{\sin 2\phi} & \, = \, 
- \, {\cal C} \, \Bigg[ \frac{2\big( \vec{h} \cdot \vec{k}_{aT} \big) 
                          \big( \vec{h} \cdot \vec{k}_{bT} \big)
                         - \vec{k}_{aT} \cdot \vec{k}_{bT}} {M_{a} M_{b}} \, 
                  h_{1}^{\perp} \, \bar{h}_{1L}^{\perp} \Bigg] \,,
\displaybreak[0] \\ \label{e:tu_1}
F_{TU}^{1} & \, = \,  
- \, {\cal C} \, \Bigg[ \frac{\vec{h} \cdot \vec{k}_{aT}} {M_{a}} \, 
                  f_{1T}^{\perp} \, \bar{f}_{1} \Bigg] \,,
\displaybreak[0] \\ \label{e:tu_2}
F_{TU}^{\sin (2\phi - \phi_a)} & \, = \, 
{\cal C} \, \Bigg[ \frac{\vec{h} \cdot \vec{k}_{bT}} {M_{b}} \,
                  h_{1} \, \bar{h}_{1}^{\perp} \Bigg] \,,
\displaybreak[0] \\
F_{TU}^{\sin (2\phi + \phi_a)} & \, = \, 
{\cal C} \, \Bigg[ \frac{2\big( \vec{h} \cdot \vec{k}_{aT} \big)
                    \big[2 \big( \vec{h} \cdot\vec{k}_{aT} \big)
                         \big( \vec{h} \cdot \vec{k}_{bT} \big)
                        -\vec{k}_{aT} \cdot \vec{k}_{bT} \big]
                   - \vec{k}_{aT}^{2} \big( \vec{h} \cdot \vec{k}_{bT} \big)}
                  {2 M_{a}^{2} M_{b}} \, 
                  h_{1T}^{\perp} \, \bar{h}_{1}^{\perp} \Bigg] \,,
\displaybreak[0] \\ \label{e:ut_1}
F_{UT}^{1} & \, = \,  
{\cal C} \, \Bigg[ \frac{\vec{h} \cdot \vec{k}_{bT}} {M_{b}} \, 
                  f_{1} \, \bar{f}_{1T}^{\perp} \Bigg] \,,
\displaybreak[0] \\ \label{e:ut_2}
F_{UT}^{\sin (2\phi - \phi_b)} & \, = \, 
- \, {\cal C} \, \Bigg[ \frac{\vec{h} \cdot \vec{k}_{aT}} {M_{a}} \,
                  h_{1}^{\perp} \, \bar{h}_{1} \Bigg] \,,
\displaybreak[0] \\
F_{UT}^{\sin (2\phi + \phi_b)} & \, = \, 
- \, {\cal C} \, \Bigg[ \frac{2\big( \vec{h} \cdot \vec{k}_{bT} \big)
                    \big[2 \big( \vec{h} \cdot\vec{k}_{aT} \big)
                         \big( \vec{h} \cdot \vec{k}_{bT} \big)
                        -\vec{k}_{aT} \cdot \vec{k}_{bT} \big]
                   - \vec{k}_{bT}^{2} \big( \vec{h} \cdot \vec{k}_{aT} \big)}
                  {2 M_{a} M_{b}^{2}} \, 
                  h_{1}^{\perp} \, \bar{h}_{1T}^{\perp} \Bigg] \,,
\displaybreak[0] \\
F_{LL}^{1} & \, = \,  
- \, {\cal C} \, \big[ g_{1L} \, \bar{g}_{1L} \big] \,, 
\vphantom{\Bigg[} 
\displaybreak[0] \\
F_{LL}^{\cos 2\phi} & \, = \, 
{\cal C} \, \Bigg[ \frac{2\big( \vec{h} \cdot \vec{k}_{aT} \big) 
                          \big( \vec{h} \cdot \vec{k}_{bT} \big)
                         - \vec{k}_{aT} \cdot \vec{k}_{bT}} {M_{a} M_{b}} \, 
                  h_{1L}^{\perp} \bar{h}_{1L}^{\perp} \Bigg] \,,
\displaybreak[0] \\
F_{LT}^{1} & \, = \,  
- \, {\cal C} \, \Bigg[ \frac{\vec{h} \cdot \vec{k}_{bT}} {M_{b}} \, 
                  g_{1L} \, \bar{g}_{1T} \Bigg] \,,
\displaybreak[0] \\
F_{LT}^{\cos (2\phi - \phi_b)} & \, = \, 
{\cal C} \, \Bigg[ \frac{\vec{h} \cdot \vec{k}_{aT}} {M_{a}} \,
                  h_{1L}^{\perp} \, \bar{h}_{1} \Bigg] \,,
\displaybreak[0] \\
F_{LT}^{\cos (2\phi + \phi_b)} & \, = \, 
{\cal C} \, \Bigg[ \frac{2\big( \vec{h} \cdot \vec{k}_{bT} \big)
                    \big[2 \big( \vec{h} \cdot\vec{k}_{aT} \big)
                         \big( \vec{h} \cdot \vec{k}_{bT} \big)
                        -\vec{k}_{aT} \cdot \vec{k}_{bT} \big]
                   - \vec{k}_{bT}^{2} \big( \vec{h} \cdot \vec{k}_{aT} \big)}
                  {2 M_{a} M_{b}^{2}} \, 
                  h_{1L}^{\perp} \, \bar{h}_{1T}^{\perp} \Bigg] \,,
\displaybreak[0] \\
F_{TL}^{1} & \, = \,  
- \, {\cal C} \, \Bigg[ \frac{\vec{h} \cdot \vec{k}_{aT}} {M_{a}} \, 
                  g_{1T} \, \bar{g}_{1L} \Bigg] \,,
\displaybreak[0] \\
F_{TL}^{\cos (2\phi - \phi_a)} & \, = \, 
{\cal C} \, \Bigg[ \frac{\vec{h} \cdot \vec{k}_{bT}} {M_{b}} \,
                  h_{1} \, \bar{h}_{1L}^{\perp} \Bigg] \,,
\displaybreak[0] \\
F_{TL}^{\cos (2\phi + \phi_a)} & \, = \, 
{\cal C} \, \Bigg[ \frac{2\big( \vec{h} \cdot \vec{k}_{aT} \big)
                    \big[2 \big( \vec{h} \cdot\vec{k}_{aT} \big)
                         \big( \vec{h} \cdot \vec{k}_{bT} \big)
                        -\vec{k}_{aT} \cdot \vec{k}_{bT} \big]
                   - \vec{k}_{aT}^{2} \big( \vec{h} \cdot \vec{k}_{bT} \big)}
                  {2 M_{a}^{2} M_{b}} \, 
                  h_{1T}^{\perp} \, \bar{h}_{1L}^{\perp} \Bigg] \,,
\displaybreak[0] \\ \label{e:tt_1}
F_{TT}^{1} & \, = \,  
{\cal C} \, \Bigg[ \frac{2\big( \vec{h} \cdot \vec{k}_{aT} \big) 
                          \big( \vec{h} \cdot \vec{k}_{bT} \big)
                         - \vec{k}_{aT} \cdot \vec{k}_{bT}} {2 M_{a} M_{b}} \, 
               \Big( f_{1T}^{\perp} \, \bar{f}_{1T}^{\perp}
                   - g_{1T} \, \bar{g}_{1T} \Big) \Bigg] \,,
\displaybreak[0] \\ \label{e:tt_2}
\bar{F}_{TT}^{1} & \, = \, 
- \, {\cal C} \, \Bigg[ \frac{\vec{k}_{aT} \cdot \vec{k}_{bT}} {2 M_{a} M_{b}} \, 
               \Big( f_{1T}^{\perp} \, \bar{f}_{1T}^{\perp}
                   + g_{1T} \, \bar{g}_{1T} \Big) \Bigg] \,,
\\
F_{TT}^{\cos (2\phi - \phi_a - \phi_b)} & \, = \,  
{\cal C} \, \big[ h_{1} \, \bar{h}_{1} \big] \,, 
\vphantom{\Bigg[} 
\displaybreak[0] \\
F_{TT}^{\cos (2\phi - \phi_a + \phi_b)} & \, = \,  
{\cal C} \, \Bigg[ \frac{2 \big(\vec{h} \cdot \vec{k}_{bT} \big)^{2} - \vec{k}_{bT}^{2}}
                        {2M_{b}^{2}} \, h_{1} \, \bar{h}_{1T}^{\perp} \bigg] \,,
\displaybreak[0] \\
F_{TT}^{\cos (2\phi + \phi_a - \phi_b)} & \, = \,  
{\cal C} \, \Bigg[ \frac{2 \big(\vec{h} \cdot \vec{k}_{aT} \big)^{2} - \vec{k}_{aT}^{2}}
                        {2M_{a}^{2}} \, h_{1T}^{\perp} \, \bar{h}_{1} \bigg] \,,
\\ \label{e:tt_6}
F_{TT}^{\cos (2\phi + \phi_a + \phi_b)} & \, = \, 
{\cal C} \, \Bigg[ \Bigg( \frac{4 \big( \vec{h} \cdot \vec{k}_{aT} \big) 
                                  \big( \vec{h} \cdot \vec{k}_{bT} \big)
                    \big[2 \big( \vec{h} \cdot \vec{k}_{aT} \big)
                           \big( \vec{h} \cdot \vec{k}_{bT})
                         - \vec{k}_{aT}\cdot\vec{k}_{bT} \big]} 
                               {4 M_{a}^{2} M_{b}^{2}}
\nonumber \\
& \hspace{1.2cm} + \, \frac{\vec{k}_{aT}^{2} \vec{k}_{bT}^{2} 
         - 2 \vec{k}_{aT}^{2} \big( \vec{h} \cdot \vec{k}_{bT} \big)^{2}
         - 2 \vec{k}_{bT}^{2} \big( \vec{h} \cdot \vec{k}_{aT} \big)^{2}}
                               {4 M_{a}^{2} M_{b}^{2}} \Bigg) \,
                     h_{1T}^{\perp} \, \bar{h}_{1T}^{\perp} \Bigg] \,. 
\end{align}
We close this section with a number of comments.
\begin{itemize}
\item The structure functions depend on the variables $(x_a, x_b, q_T)$. 
 Instead of using $q_T$ one may also work with the transverse momentum of one
 of the hadrons in the CS-frame.
\item One finds nonzero contributions for 24 out of the 48 structure functions
 defined in Eq.~(\ref{e:ang_dist}).
 This also means that exactly half of the structure functions are of subleading 
 twist for the kinematical region $q_T \ll q$ we are interested in here.
\item The leading twist parton model calculation containing T-even effects 
 was first carried out in Ref.~\cite{tangerman_94a}, while T-odd effects were 
 investigated in~\cite{boer_99}.
 We obtain the same number of nonzero structures identified in those articles, 
 though we do not agree with certain angular dependences given in~\cite{boer_99}.
\item Our results are for the structure functions in Eq.~(\ref{e:ang_dist}) with
 the lepton angles understood in the CS-frame, and the components of the hadron 
 spin vectors in the {\it cm}-frame.
 Note that the expressions would be exactly the same for structure functions 
 defined in the Gottfried-Jackson frame, because differences between those two 
 dilepton rest frames are only of ${\cal O}(q_T/q)$. 
\item For identical hadrons in the initial state the results in 
 Eqs.~(\ref{e:uu_1})--(\ref{e:tt_6}) satisfy the model-independent constraints
 listed in~(\ref{e:sym_f}) and~(\ref{e:rel_f}).
 In particular, we point out that the parton model result
 \begin{equation}
 F_{TT}^{\cos (2\phi + \phi_a - \phi_b)}(x_b,x_a) = 
 F_{TT}^{\cos (2\phi - \phi_a + \phi_b)}(x_a,x_b)  
 \end{equation}
 has a model-independent status.
 It is worthwhile to mention that, by means of charge conjugation, in the case of 
 proton-antiproton DY one also finds symmetries for structure functions 
 (like $F_{UU}(x_a,x_b,q_T) = F_{UU}(x_b,x_a,q_T)$), and relations between various 
 structure functions.
 In particular, when studying single spin effects one can obtain the same 
 information by either polarizing the proton or the antiproton. 
\item If the cross section is integrated upon $q_T$ only three structure 
 functions ($F_{UU}^{1}$, $F_{LL}^{1}$, $F_{TT}^{\cos (2\phi - \phi_a - \phi_b)}$) 
 survive.
 Neglecting hadron masses one obtains 
 \begin{align} \label{e:xs_qtint}
 & \frac{d \sigma}{d x_a \, d x_b \, d\Omega} = 
 \frac{s}{2} \, \frac{d \sigma}{dq^+ \, dq^- \, d\Omega} 
 \nonumber \\
 & \hspace{0.5cm}
  = \frac{\alpha_{em}^2}{12 \, q^2} \, \Big \{ 
  (1 + \cos^2 \theta) \, \sum_q \, e_q^2 \, 
       \Big( f_1^q(x_a) \, f_1^{\bar{q}}(x_b) + f_1^{\bar{q}}(x_a) \, f_1^q(x_b) \Big)
 \nonumber \\
 & \hspace{2.05cm}
  - S_{aL} \, S_{bL} \,
  (1 + \cos^2 \theta) \, \sum_q \, e_q^2 \, 
       \Big( g_1^q(x_a) \, g_1^{\bar{q}}(x_b) + g_1^{\bar{q}}(x_a) \, g_1^q(x_b) \Big)
 \nonumber \\
 &  \hspace{2.05cm}
  + |\vec{S}_{aT}| \, |\vec{S}_{bT}| \, 
  \sin^2 \theta \, \cos (2\phi - \phi_a - \phi_b) \, \sum_q \, e_q^2 \,
       \Big( h_1^q(x_a) \, h_1^{\bar{q}}(x_b) + h_1^{\bar{q}}(x_a) \, h_1^q(x_b) \Big)
 \Big \} \,. 
 \end{align}
 Further integration upon the solid angle $\Omega$ provides
 \begin{align} \label{e:xs_omegaint}
 \frac{d \sigma}{d x_a \, d x_b} 
 & = \frac{4\pi \, \alpha_{em}^2}{9 \, q^2} \, \Big \{ 
   \sum_q \, e_q^2 \, 
      \Big( f_1^q(x_a) \, f_1^{\bar{q}}(x_b) + f_1^{\bar{q}}(x_a) \, f_1^q(x_b) \Big)
 \nonumber \\
 & \hspace{1.8cm}
   - S_{aL} \, S_{bL} \, \sum_q \, e_q^2 \, 
       \Big( g_1^q(x_a) \, g_1^{\bar{q}}(x_b) + g_1^{\bar{q}}(x_a) \, g_1^q(x_b) \Big)
 \Big\} \,. 
 \end{align}
 Note that the term containing the transversity dropped out. 
\item For the $q_T$-dependent cross section all chiral-odd parton distributions 
 disappear after integrating out the azimuthal angle $\phi$.
 On the other hand, all the chiral-even effects survive this integration.
\item The large number of independent structure functions 
 --- for instance 16 for identical hadrons in the initial state --- 
 indicates the high potential of the polarized DY process for studying TMDs.
 Therefore, this process has also a certain advantage over semi-inclusive DIS 
 (if in that reaction polarization of the initial state lepton and hadron are 
 exploited) where eight leading twist structure functions 
 exist~\cite{mulders_95,bacchetta_06}, being just sufficient to map out, 
 in principle, all the eight leading twist TMDs.
\item As already pointed out in Section V data on 
 $\pi^- \, N \to \mu^- \, \mu^+ \, X$~\cite{falciano_86,guanziroli_87,conway_89}
 show a rather large $\cos 2\phi$ dependence of the unpolarized cross section 
 which cannot be explained by collinear perturbative QCD.
 However, if intrinsic transverse parton motion in the initial state is taken
 into account the Boer-Mulders function $h_1^{\perp}$ contributes to the 
 $\cos 2\phi$ term according to~(\ref{e:uu_2}) which may explain the observed 
 violation of the Lam-Tung relation~\cite{boer_99}.
 This finding stimulated a lot of phenomenological work on this 
 subject~\cite{boer_99,boer_02,lu_04,bianconi_04,lu_05,bianconi_05a,sissakian_05a,gamberg_05,sissakian_05b,lu_06,barone_06,lu_07a,reimer_07,miller_07,zhang_08,bianconi_08,sissakian_08}.
\item Of particular interest is also the transverse single spin effect given by
 $F_{TU}^{1}$ in Eq.~(\ref{e:tu_1}) or $F_{UT}^{1}$ in~(\ref{e:ut_1}).
 Both structure functions contain the Sivers parton distribution which was 
 predicted to have the opposite sign in DY as compared to semi-inclusive 
 DIS~\cite{collins_02,brodsky_02a,brodsky_02b}.
 As the sign reversal is at the core of our present understanding of transverse
 single spin asymmetries in hard scattering processes an experimental check of
 this prediction is of utmost importance.
 Theoretical work on the Sivers effect in DY can be found
 in~\cite{anselmino_02,efremov_04b,anselmino_05b,vogelsang_05,collins_05b,bianconi_05b,bianconi_06,sissakian_08}.
\item The expected sign reversal of T-odd TMDs can also be investigated through 
 the structure functions $F_{TU}^{\sin (2\phi - \phi_a)}$ in~(\ref{e:tu_2}) or 
 $F_{UT}^{\sin (2\phi - \phi_b)}$ in~(\ref{e:ut_2}) in which the Boer-Mulders 
 function enters (see also Refs.~\cite{sissakian_05a,sissakian_05b,sissakian_08}).
\item A phenomenological study of the structure functions 
 in~(\ref{e:tt_1}), (\ref{e:tt_2}) was carried out in~\cite{lu_07b}.
\end{itemize}

\section{Summary}

We have presented a formalism for dilepton production from the collision of two
polarized spin-$\tfrac{1}{2}$ particles.
To this end we have derived in a first step a general expression for the hadronic
DY tensor.
This tensor consists of 48 basis elements, and each basis tensor is multiplied
by a scalar function (structure function).
In order to ensure electromagnetic gauge invariance of the hadronic tensor we 
have made use of an elegant projection method proposed in~\cite{bardeen_68}.
In general, our treatment completes earlier work~\cite{ralston_79,pire_83}.
The double polarized case, which is the most challenging part, was studied before
only for the specific kinematical case $q_T = 0$~\cite{ralston_79}.

The result for the hadronic tensor allows one to obtain the general angular
distribution of the cross section for any reference frame.
In this work we have focussed on a dilepton rest frame where the angular 
distribution takes the most compact form and shows a high degree of symmetry.
We repeat here that the angular distribution as given in Eq.~(\ref{e:ang_dist}),
which represents a central result of our work, holds for any dilepton rest
frame.

Our analysis is supplemented by a parton model calculation of the polarized
DY reaction (see also~\cite{tangerman_94a,boer_99}).
For this part of the work we concentrated on the kinematical situation where 
the transverse momentum of the dilepton pair is much smaller than its invariant 
mass.
This region is the realm of TMDs which are currently under intense investigation
both from the experimental and the theoretical side.

We reemphasize that the polarized DY process has a high potential for studying 
TMDs which contain important information on the nonperturbative structure of 
hadrons.
Moreover, polarized dilepton measurements can provide us with a crucial and 
highly nontrivial check of QCD-factorization~\cite{collins_02}.
In addition, one can systematically study different resummation 
techniques~\cite{sterman_86,catani_89,catani_90} in an unprecedented way.
Consequently, there is sufficient reason for looking forward to the first 
polarized DY data.

\begin{acknowledgments}
This work has been partially supported by the Verbundforschung 
``Hadronen und Kerne'' of the BMBF and by the 
Deutsche Forschungsgemeinschaft (DFG).
\\[0.3cm]
\noindent
\textbf{Notice:} Authored by Jefferson Science Associates, LLC under U.S. DOE
Contract No. DE-AC05-06OR23177.
The U.S. Government retains a non-exclusive, paid-up, irrevocable, world-wide 
license to publish or reproduce this manuscript for U.S. Government purposes.
\end{acknowledgments}


\end{document}